\pdfoutput=1
\documentclass[12pt,a4paper]{article}

\usepackage{ifthen} 
\newboolean{pdflatex}
\setboolean{pdflatex}{true} 

\newboolean{articletitles}
\setboolean{articletitles}{true} 

\newboolean{uprightparticles}
\setboolean{uprightparticles}{false} 

\def\paperauthors{LHCb collaboration} 
\def\paperasciititle{Constraints on the photon polarisation in bsgamma transitions using Bs to phi ee decays}  
\def\papertitle{Constraints on the photon polarisation in $b \to s \gamma$ transitions using $B_s^0 \rightarrow \phi e^+e^-$ decays} 
\def\paperkeywords{{High Energy Physics}, {LHCb}}
\def\papercopyright{\the\year\ CERN for the benefit of the LHCb collaboration}
\def\paperlicence{CC BY 4.0 licence}
\def\paperlicenceurl{https://creativecommons.org/licenses/by/4.0/}

\newif\ifEnableSectionTOCLinks
\EnableSectionTOCLinksfalse 

\usepackage[top=1in, bottom=1.25in, left=1in, right=1in]{geometry}

\usepackage{microtype}
\usepackage{lineno}  
\usepackage{xspace} 
\usepackage{caption} 

\usepackage{graphicx}  
\usepackage{color}
\usepackage{colortbl}
\graphicspath{{./figs/}} 

\usepackage{amsmath} 
\usepackage{amssymb}
\usepackage{amsfonts}
\usepackage{upgreek} 

\usepackage{floatrow}
\newcommand*\patchAmsMathEnvironmentForLineno[1]{%
\expandafter\let\csname old#1\expandafter\endcsname\csname #1\endcsname
\expandafter\let\csname oldend#1\expandafter\endcsname\csname
end#1\endcsname
 \renewenvironment{#1}%
   {\linenomath\csname old#1\endcsname}%
   {\csname oldend#1\endcsname\endlinenomath}%
}
\newcommand*\patchBothAmsMathEnvironmentsForLineno[1]{%
  \patchAmsMathEnvironmentForLineno{#1}%
  \patchAmsMathEnvironmentForLineno{#1*}%
}
\AtBeginDocument{%
\patchBothAmsMathEnvironmentsForLineno{equation}%
\patchBothAmsMathEnvironmentsForLineno{align}%
\patchBothAmsMathEnvironmentsForLineno{flalign}%
\patchBothAmsMathEnvironmentsForLineno{alignat}%
\patchBothAmsMathEnvironmentsForLineno{gather}%
\patchBothAmsMathEnvironmentsForLineno{multline}%
\patchBothAmsMathEnvironmentsForLineno{eqnarray}%
}

\usepackage{hyperxmp}

\usepackage[pdftex,
            pdfauthor={\paperauthors},
            pdftitle={\paperasciititle},
            pdfkeywords={\paperkeywords},
            pdfcopyright={Copyright (C) \papercopyright},
            pdflicenseurl={\paperlicenceurl}]{hyperref}

\usepackage[colorinlistoftodos,textsize=scriptsize]{todonotes}

\usepackage[bottom,flushmargin,hang,multiple]{footmisc}

\usepackage[all]{hypcap} 

\usepackage{xspace} 
\usepackage{upgreek}

\def\lhcb   {\mbox{LHCb}\xspace}

\def\babar  {\mbox{BaBar}\xspace}

\def\MagUp {\mbox{\em Mag\kern -0.05em Up}\xspace}

\ifthenelse{\boolean{uprightparticles}}%
{
 
 \def\Pgamma      {\ensuremath{\upgamma}\xspace}

 \def\Ppi         {\ensuremath{\uppi}\xspace}

 \def\PDelta      {\ensuremath{\Delta}\xspace}                 
 \def\PXi         {\ensuremath{\Xi}\xspace}                 
 \def\PLambda     {\ensuremath{\Lambda}\xspace}                 
 \def\PSigma      {\ensuremath{\Sigma}\xspace}                 
 \def\POmega      {\ensuremath{\Omega}\xspace}                 
 \def\PUpsilon    {\ensuremath{\Upsilon}\xspace}
 \let\oldPi\Pi
 \def\PPi         {\ensuremath{\oldPi}\xspace}

 \def\PB      {\ensuremath{\mathrm{B}}\xspace}                 
 \def\PD      {\ensuremath{\mathrm{D}}\xspace}                 

 \def\PK      {\ensuremath{\mathrm{K}}\xspace}                 
 \def\Pb      {\ensuremath{\mathrm{b}}\xspace}

 \def\Pe      {\ensuremath{\mathrm{e}}\xspace}                 
 \def\Ps      {\ensuremath{\mathrm{s}}\xspace}

 \def\thebaroffset{0.0em}
}
{
 
 \def\Pgamma      {\ensuremath{\gamma}\xspace}

 \def\Ppi         {\ensuremath{\pi}\xspace}

 \mathchardef\PDelta="7101
 \mathchardef\PXi="7104
 \mathchardef\PLambda="7103
 \mathchardef\PSigma="7106
 \mathchardef\POmega="710A
 \mathchardef\PUpsilon="7107
 \mathchardef\PPi="7105
 \def\PB      {\ensuremath{B}\xspace}                 
 \def\PD      {\ensuremath{D}\xspace}                 

 \def\PK      {\ensuremath{K}\xspace}                 
 \def\Pb      {\ensuremath{b}\xspace}

 \def\Pe      {\ensuremath{e}\xspace}                 
 \def\Ps      {\ensuremath{s}\xspace}

 \def\thebaroffset{0.18em}
}
\newcommand{\offsetoverline}[2][\thebaroffset]{\kern #1\overline{\kern -#1 #2}}%

\makeatletter
\ifcase \@ptsize \relax
  \newcommand{\miniscule}{\@setfontsize\miniscule{4}{5}}
\or
  \newcommand{\miniscule}{\@setfontsize\miniscule{5}{6}}
\or
  \newcommand{\miniscule}{\@setfontsize\miniscule{5}{6}}
\fi
\makeatother

\DeclareRobustCommand{\optbar}[1]{\shortstack{{\miniscule (\rule[.5ex]{1.25em}{.18mm})}
  \\ [-.7ex] $#1$}}

\def\ep         {{\ensuremath{\Pe^+}}\xspace}

\def\epem       {{\ensuremath{\Pe^+\Pe^-}}\xspace}

\def\g      {{\ensuremath{\Pgamma}}\xspace}

\def\squark    {{\ensuremath{\Ps}}\xspace}

\def\bquark    {{\ensuremath{\Pb}}\xspace}

\def\pion   {{\ensuremath{\Ppi}}\xspace}
\def\piz    {{\ensuremath{\pion^0}}\xspace}

\def\kaon    {{\ensuremath{\PK}}\xspace}

\def\KorKbar {\kern \thebaroffset\optbar{\kern -\thebaroffset \PK}{}\xspace}

\def\Kp      {{\ensuremath{\kaon^+}}\xspace}
\def\Km      {{\ensuremath{\kaon^-}}\xspace}

\def\Kstarz  {{\ensuremath{\kaon^{*0}}}\xspace}

\def\D       {{\ensuremath{\PD}}\xspace}

\def\DorDbar {\kern \thebaroffset\optbar{\kern -\thebaroffset \PD}\xspace}

\def\Dp      {{\ensuremath{\D^+}}\xspace}
\def\Dm      {{\ensuremath{\D^-}}\xspace}

\def\DpDm    {\ensuremath{\Dp {\kern -0.16em \Dm}}\xspace}

\def\Dsm     {{\ensuremath{\D^-_\squark}}\xspace}

\def\Dssm    {{\ensuremath{\D^{*-}_\squark}}\xspace}

\def\B       {{\ensuremath{\PB}}\xspace}
\def\Bbar    {{\ensuremath{\offsetoverline{\PB}}}\xspace}

\def\BorBbar {\kern \thebaroffset\optbar{\kern -\thebaroffset \PB}\xspace}

\def\Bd      {{\ensuremath{\B^0}}\xspace}

\def\BdorBdbar {\kern \thebaroffset\optbar{\kern -\thebaroffset \Bd}\xspace}

\def\Bs      {{\ensuremath{\B^0_\squark}}\xspace}
\def\Bsb     {{\ensuremath{\Bbar{}^0_\squark}}\xspace}
\def\BsorBsbar {\kern \thebaroffset\optbar{\kern -\thebaroffset \Bs}\xspace}

\def\Y#1S{\ensuremath{\PUpsilon{(#1S)}}\xspace}

\def\Lz          {{\ensuremath{\PLambda}}\xspace}

\def\LorLbar     {\kern \thebaroffset\optbar{\kern -\thebaroffset \PLambda}\xspace}

\def\Lb           {{\ensuremath{\Lz^0_\bquark}}\xspace}

\newcommand{\decay}[2]{\ensuremath{#1\!\to #2}\xspace} 

\def\to                 {\ensuremath{\rightarrow}\xspace}

\def\qsq       {{\ensuremath{q^2}}\xspace}

\def\CP                {{\ensuremath{C\!P}}\xspace}

\def\BsPhiGam     {\decay{\Bs}{\phi \g}}

\def\BsPhiEE     {\decay{\Bs}{\phi \epem}}
\def\BsPhiPiz     {\decay{\Bs}{\phi \piz}}
\def\BsPhiEta     {\decay{\Bs}{\phi \eta}}
\def\BdKstee  {\decay{\Bd}{\Kstarz\epem}}

\def\bTosee    {\decay{\bquark}{\squark \epem}}

\def\FL       {\ensuremath{F_{\mathrm{L}}}\xspace}
\def\AT#1     {\ensuremath{A_{\mathrm{T}}^{#1}}\xspace}           
\def\btosgam  {\decay{\bquark}{\squark \g}}

\def\ctl       {\ensuremath{\cos{\theta_L}}\xspace}
\def\ctk       {\ensuremath{\cos{\theta_K}}\xspace}

\def\C#1      {\ensuremath{\mathcal{C}_{#1}}\xspace}                       
\def\Cp#1     {\ensuremath{\mathcal{C}_{#1}^{'}}\xspace}                    
\def\Ceff#1   {\ensuremath{\mathcal{C}_{#1}^{\mathrm{(eff)}}}\xspace}        
\def\Cpeff#1  {\ensuremath{\mathcal{C}_{#1}^{'\mathrm{(eff)}}}\xspace}       
\def\Ope#1    {\ensuremath{\mathcal{O}_{#1}}\xspace}                       
\def\Opep#1   {\ensuremath{\mathcal{O}_{#1}^{'}}\xspace}                    

\newcommand{\aunit}[1]{\ensuremath{\text{\,#1}}}       

\newcommand{\tev}{\aunit{Te\kern -0.1em V}\xspace}
\newcommand{\gev}{\aunit{Ge\kern -0.1em V}\xspace}
\newcommand{\mev}{\aunit{Me\kern -0.1em V}\xspace}
\newcommand{\kev}{\aunit{ke\kern -0.1em V}\xspace}
\newcommand{\ev}{\aunit{e\kern -0.1em V}\xspace}
 
\newcommand{\mevc}{\ensuremath{\aunit{Me\kern -0.1em V\!/}c}\xspace}
\newcommand{\gevc}{\ensuremath{\aunit{Ge\kern -0.1em V\!/}c}\xspace}
\newcommand{\mevcc}{\ensuremath{\aunit{Me\kern -0.1em V\!/}c^2}\xspace}
\newcommand{\gevcc}{\ensuremath{\aunit{Ge\kern -0.1em V\!/}c^2}\xspace}
\newcommand{\gevgevcccc}{\ensuremath{\gev^2\!/c^4}\xspace} 

\def\fb   {\ensuremath{\aunit{fb}}\xspace}
\def\invfb   {\ensuremath{\fb^{-1}}\xspace}

\newcommand{\chisqip}{\ensuremath{\chi^2_{\text{IP}}}\xspace}

\def\deriv {\ensuremath{\mathrm{d}}}

\def\gsim{{~\raise.15em\hbox{$>$}\kern-.85em
          \lower.35em\hbox{$\sim$}~}\xspace}
\def\lsim{{~\raise.15em\hbox{$<$}\kern-.85em
          \lower.35em\hbox{$\sim$}~}\xspace}

\newcommand{\Real}{\ensuremath{\mathcal{R}e}\xspace}
\newcommand{\Imag}{\ensuremath{\mathcal{I}m}\xspace}

\def\pt         {\ensuremath{p_{\mathrm{T}}}\xspace}

\def\geant      {\mbox{\textsc{Geant4}}\xspace}

\def\photos     {\mbox{\textsc{Photos}}\xspace}

\def\pythia     {\mbox{\textsc{Pythia}}\xspace}

\def\tell1  {TELL1\xspace}
\def\ukl1   {UKL1\xspace}

\newcommand{\ie}{\mbox{\itshape i.e.}\xspace}

\newcommand{\lhcborcid}[1]{\href{https://orcid.org/#1}{\hspace*{0.1em}\raisebox{-0.45ex}{\includegraphics[width=1em]{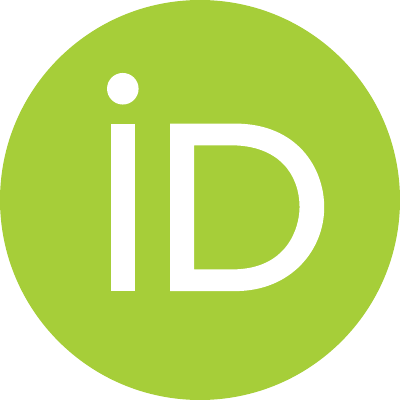}}}}

\hypersetup{
  colorlinks   = true, 
  urlcolor     = blue, 
  linkcolor    = blue, 
  citecolor    = red   
}

\ifEnableSectionTOCLinks
    \usepackage[explicit]{titlesec} 
    
    \let\oldcontentsline\contentsline
    \renewcommand

    \titleformat{\section}{\normalfont\Large\bf}{\hyperlink{tocsection.\thesection}{{\thesection} \parbox[t]{\dimexpr\textwidth-1pc}{#1}}}{1pc}{}

    \titleformat{\subsection}{\normalfont\bf}{\hyperlink{tocsubsection.\thesubsection}{{\thesubsection} \parbox[t]{\dimexpr\textwidth-1pc}{#1}}}{1pc}{}

    \titleformat{name=\section,numberless}[display]{}{}{0pt}{\normalfont\Huge\bfseries #1}
\fi

\usepackage{cite} 
\usepackage{mciteplus}

\usepackage{longtable}

\begin{document}

\renewcommand{\thefootnote}{\fnsymbol{footnote}}
\setcounter{footnote}{1}

\begin{titlepage}
\pagenumbering{roman}

\vspace*{-1.5cm}
\centerline{\large EUROPEAN ORGANIZATION FOR NUCLEAR RESEARCH (CERN)}
\vspace*{1.5cm}
\noindent
\begin{tabular*}{\linewidth}{lc@{\extracolsep{\fill}}r@{\extracolsep{0pt}}}
\ifthenelse{\boolean{pdflatex}}
{\vspace*{-1.5cm}\mbox{\!\!\!\includegraphics[width=.14\textwidth]{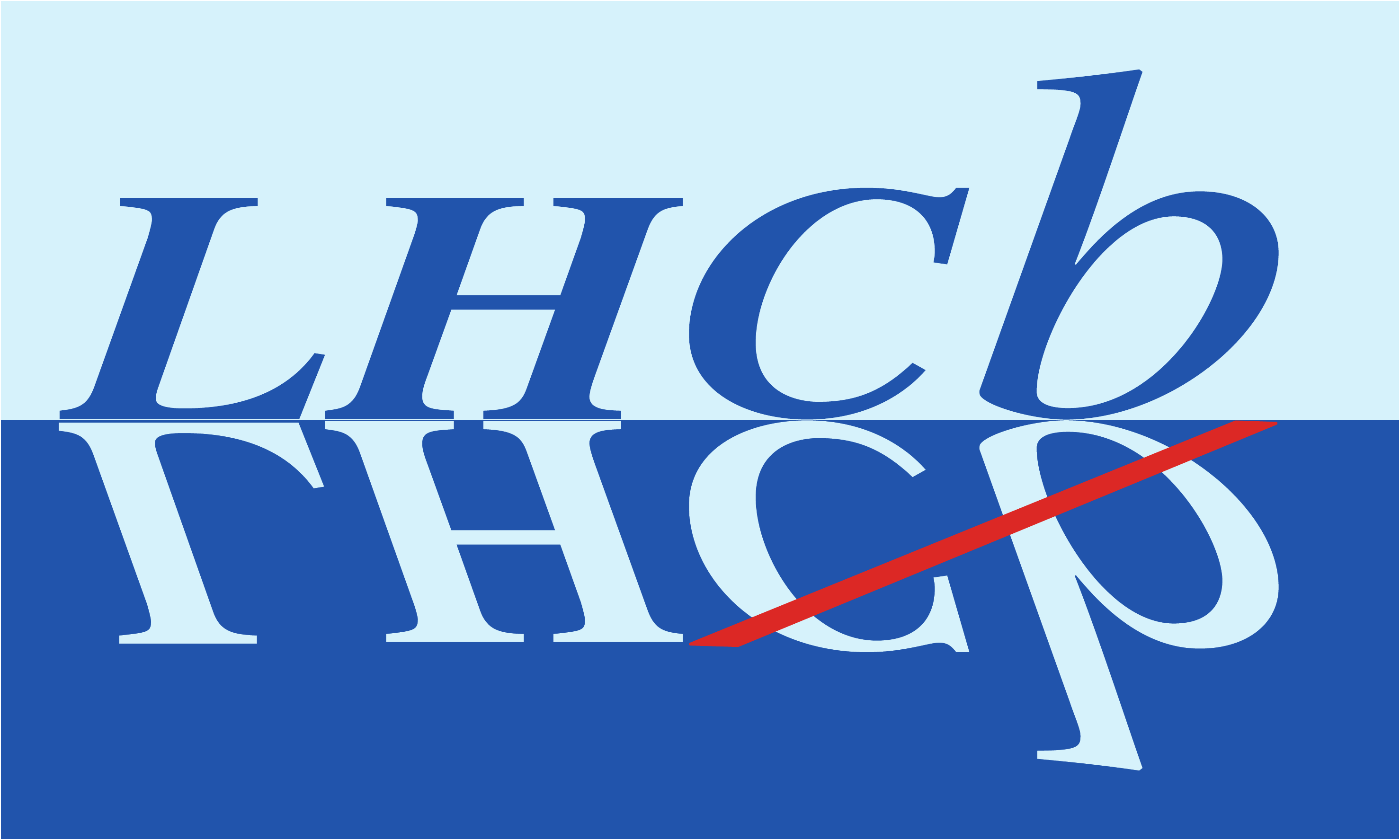}} & &}%
{\vspace*{-1.2cm}\mbox{\!\!\!\includegraphics[width=.12\textwidth]{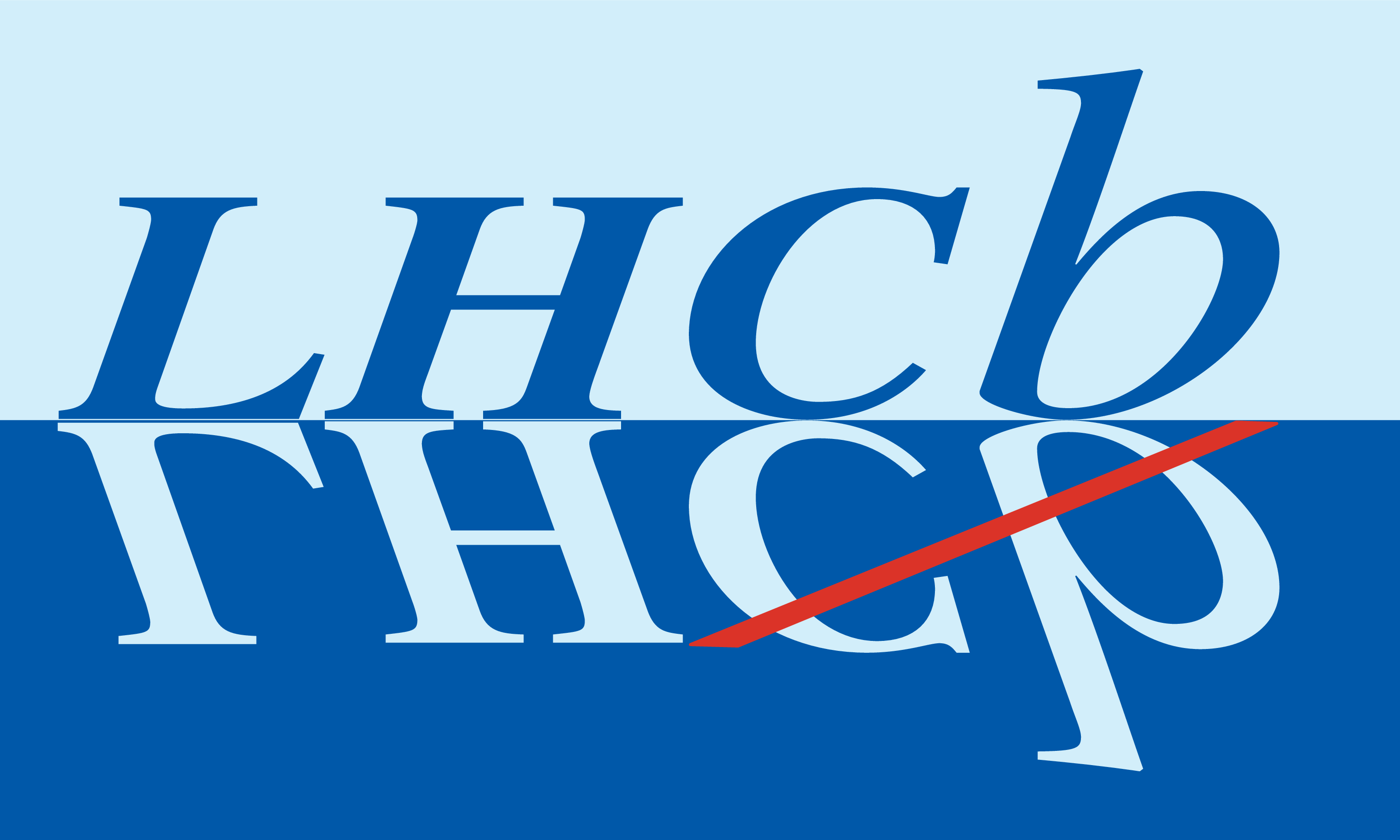}} & &}%
\\
 & & CERN-EP-2024-276 \\  
 & & LHCb-PAPER-2024-030 \\  
 & & March 7, 2025 \\
 & & \\
\end{tabular*}

\vspace*{4.0cm}

{\normalfont\bfseries\boldmath\huge
\begin{center}
  \papertitle 
\end{center}
}

\vspace*{1.5cm}

\begin{center}
\paperauthors\footnote{Authors are listed at the end of this paper.}
\end{center}

\vspace{\fill}

\begin{abstract}
  \noindent
  An angular analysis of the $B_s^0 \rightarrow \phi e^+e^-$ decay is performed using the  proton-proton collision dataset collected between 2011 and 2018  by the LHCb experiment, corresponding to an integrated luminosity of $9\,{\rm fb}^{-1}$ at centre-of-mass energies of 7, 8 and $13\,{\rm TeV}$. The analysis is performed in the very low dielectron invariant mass-squared region between $0.0009$ and $0.2615\,{\rm GeV}^2\!/c^4$.
 The longitudinal polarisation fraction of the $\phi$ meson is measured to be less than 11.5\% at 90\% confidence level. The $A_{\mathrm{T}}^{\mathcal{R}e C\!P}$ observable, which is related to the lepton forward-backward asymmetry, is measured to be $0.116 \pm 0.155 \pm 0.006$, where the first uncertainty is statistical and the second systematic.  The transverse asymmetries, $A_{\mathrm{T}}^{(2)}$ and $A_{\mathrm{T}}^{\mathcal{I}m C\!P}$ , which are sensitive to the virtual photon polarisation, are found to be $-0.045 \pm 0.235 \pm 0.014$ and $0.002 \pm 0.247 \pm 0.016$, respectively. The results are consistent with Standard Model predictions. 

\end{abstract}

\vspace*{1.5cm}

\begin{center}
  Published in 
  JHEP {\bf 03} (2025) 047
\end{center}

\vspace{\fill}

{\footnotesize 
\centerline{\copyright~\papercopyright. \href{\paperlicenceurl}{\paperlicence}.}}
\vspace*{2mm}

\end{titlepage}


\newpage
\setcounter{page}{2}
\mbox{~}
\newpage


\renewcommand{\thefootnote}{\arabic{footnote}}
\setcounter{footnote}{0}

\cleardoublepage


\pagestyle{plain} 
\setcounter{page}{1}
\pagenumbering{arabic}


\section{Introduction}
\label{sec:Introduction}
In the Standard Model of particle physics (SM), processes involving \btosgam transitions are suppressed as they occur through flavour-changing neutral currents, mediated by loop processes involving electroweak penguin diagrams. The study of these processes provides insight into physics beyond the SM (BSM) due to potential deviations that may appear\cite{Atwood:1997zr, Everett:2001yy, Grinstein:2004uu, Foster:2006ze, Lunghi:2006hc, Becirevic:2012dx}. In the SM, the electroweak charged current has chiral interactions, coupling only to left-handed quarks. As a result, in \btosgam transitions, the photons are predominantly left-handed, with a small right-handed contribution that has a relative amplitude proportional to the ratio of the \squark- to \bquark-quark masses. The presence of a significant right-handed polarisation would indicate a clear evidence of BSM physics.

The  \babar \cite{Aubert_2008,Lees_2012,Lees_2012_2}, Belle \cite{Saito_2015,bellecollaboration2016measurement,Ushiroda_2006,Belle:2024mml}, Belle II \cite{Belle-II:2024uzp} and LHCb \cite{LHCb-PAPER-2019-015, LHCb-PAPER-2021-030} collaborations have tested the photon polarisation
in \btosgam transitions through measurements of the inclusive $B \to X_s\gamma$ branching fraction, \CP asymmetries in radiative $\Bd$ and $\Bs$ decays, and by conducting an angular analysis of radiative $\Lb$ decays. The strongest constraint to date comes from an angular analysis of \BdKstee decays\footnote{Throughout this paper, \Kstarz is used to refer to the $\Kstarz(892)$ resonance and the inclusion of charge-conjugate processes is implied unless otherwise stated.} by the \lhcb experiment\cite{LHCb-PAPER-2020-020}, isolating the dominant \btosgam contribution by restricting the dilepton invariant mass-squared, \qsq, to small values.
The \bTosee transition is particularly sensitive to the photon pole due to the smallness of the electron mass.

This paper presents the first angular analysis of the \BsPhiEE decay\footnote{The shorthand $\phi$ refers to the $\phi(1020)$ meson throughout this paper.} in the very low $\qsq$ region located between $0.0009$ and $0.2615\,\gevgevcccc$, chosen to be identical to the kinematical region of Ref.~\cite{LHCb-PAPER-2020-020}. The $\phi$ meson is reconstructed via its \decay{\phi}{\Kp \Km} decay. This analysis uses the data collected by the LHCb experiment between 2011 and 2018.
The \BsPhiEE channel is experimentally cleaner than \BdKstee as partially reconstructed background contributions from higher hadronic resonances are suppressed to a negligible level. In addition, the narrow width of the $\phi$ meson reduces sources of experimental backgrounds. Despite the lower production rate of \Bs mesons compared to $\Bd$ mesons and the smaller branching fraction of the hadronic resonance for the \Bs channel, the analysis of the \BsPhiEE decay in the very low \qsq region provides valuable and complementary insights.

The \BsPhiEE decay is described across the entire $\qsq$ spectrum by the left (right)-handed Wilson coefficients $C^{(')}_7$, $C^{(')}_9$, and $C^{(')}_{10}$~\cite{Bharucha_2016,Ball2007}. These coefficients carry crucial information regarding short-distance effects and exhibit sensitivity to BSM physics.
The very low $\qsq$ region, given the association of this decay with both left- and right-handed electromagnetic operators $O_7$ and $O^{'}_7$, is fundamental for the determination of the $C_7$ and $C^{'}_7$ Wilson coefficients, respectively.

The \BsPhiEE differential decay rate can be written as a function of \qsq\ and three angular variables referred to as \ctl, \ctk and $\varphi$. The angle $\theta_{L}$ ($\theta_{K}$) is defined between the direction of the $e^-$ (\Km) and the direction opposite that of the \Bs flight direction in the $\epem (K^+K^-)$ rest frame, while the angle $ \varphi $  is defined between the plane containing the two leptons and that containing the two final-state kaons in the $\Bs$ rest frame. As the flavour of the $\Bs$ meson at decay cannot be determined from the flavour-symmetric final state, the same angular definition is used for both $\Bs$ and $\Bsb$ decays, later referred to as ``untagged". As in Ref.~\cite{LHCb-PAPER-2014-066}, the angle $\varphi$ is transformed such that $\Tilde{\varphi} = \varphi + \pi$ if $\varphi < 0$. This transformation leads to a cancellation of the terms containing $\sin\varphi$ and $\cos\varphi$, consequently simplifying the angular expression without any loss of sensitivity to the remaining physical observables.
Ignoring any $K^+K^-$ S-wave contribution which should be negligible at low-$\qsq$ values due to the photon-pole dominance, and in the limit of massless leptons, the normalised, untagged, time-integrated, \CP-averaged angular differential decay rate is~\cite{Descotes-Genon:2022qce} \begin{equation}
    \begin{aligned}
    \frac{1}{\deriv(\Gamma+\Bar{\Gamma})/\deriv\qsq}\dfrac{\deriv^3(\Gamma+\Bar{\Gamma})}{\deriv{\cos \theta _{L}}\, \deriv{\cos  \theta _{K}}\,\deriv{\Tilde{\varphi}}}& = \frac{9}{32 \pi } \left\lbrace\frac{3}{4} \left( 1-\FL \right) \sin ^{2} \theta _{K} +\FL\cos ^{2} \theta _{K}\right. \\
  &+ \left[ \frac{1}{4} \left( 1-\FL \right) \sin ^{2} \theta _{K}-\FL\cos ^{2} \theta _{K} \right] \cos 2 \theta _{L}\\
  &+ \frac{1}{2} \left( 1-\FL \right) \AT^{ \left( 2 \right) }\sin ^{2} \theta _{K}\sin ^{2} \theta _{L}\cos 2 \Tilde{\varphi} \\
  &+  \left( 1-\FL \right) \AT^{\Real \CP}\sin ^{2} \theta _{K}\cos  \theta _{L} \\
  &\left.+ \frac{1}{2} \left( 1-\FL \right) \AT^{\Imag \CP}\sin ^{2} \theta _{K}\sin ^{2} \theta _{L}\sin 2 \Tilde{\varphi} \right \rbrace \, .
\end{aligned}
\label{eq:PhysicsPDF}
\end{equation}

The four angular observables \FL, \AT{(2)}, \AT{\Imag \CP} and \AT{\Real \CP} are combinations of the amplitudes $A_{0, \perp, \parallel}^{L,R}$ where  the indices $0, \perp, \parallel$ refer to the different polarisation states of the $\phi$ meson in the decay, and $L$ and $R$ to the left- and right-hand chirality of the dielectron system.
The observable \FL is the longitudinal polarisation fraction of the $\phi$ meson, while \AT{\Real \CP} is related to the lepton forward-backward asymmetry. The parameters \AT{(2)} and \AT{\Imag \CP} are pivotal for this analysis as they are related to the photon polarisation and thus to the Wilson coefficients $C_7^{(')}$ in the very low $q^2$ region. Indeed, following Ref.~\cite{Descotes-Genon:2022qce} and in the absence of decay-time acceptance, in the $q^2 \to 0$ limit, these observables can be expressed as
\begin{align}
\lim_{\qsq\to 0}\AT{(2)} (\qsq) &= \frac{2\bigg[ \Real [C_7]\Real[C_7^{'}] + \Imag [C_7]\Imag[C_7^{'}] + \frac{y}{2}[(\Real [C_7])^2 - (\Imag [C_7])^2]\bigg]}{(\Real [C_7])^2 + (\Imag [C_7])^2} \, , \\
\lim_{\qsq\to 0}\AT{\Imag \CP} (\qsq) &= \frac{2\bigg[ \Real [C_7]\Imag [C_7^{'}]-\Imag[C_7]\Real [C_7^{'}] - y \Real [C_7]\Imag [C_7]\bigg]}{(\Real [C_7])^2 + (\Imag [C_7])^2} \, ,
\end{align}
with $y=\frac{\Delta\Gamma_s}{2\Gamma_s}$, where $\Delta\Gamma_s$ is the difference in decay widths between the heavy and light \Bs mass eigenstates and $\Gamma_s$ the inverse of the average lifetime of the \Bs meson. The \FL and \AT{\Real \CP} parameters vanish at low values of $\qsq$ as expected from real photon interactions.

In the SM, where $C_7$ is purely real, the measurement of \AT{\Imag \CP} is identical to that obtained in the similar measurement using \BdKstee decays~\cite{LHCb-PAPER-2020-020}, while in the case of \AT{(2)} the information is slightly different due to the significant variation in decay widths between the light and the heavy mass eigenstates in the \Bs system. In the future, with larger data samples, the comparison of the \AT{(2)}  measurements obtained from \BsPhiEE and \BdKstee decays could provide additional constraints on $C_7$.

\section{The \lhcb detector and dataset}
The study reported here is based on proton--proton ($pp$) collision data collected with the \lhcb detector in the years 2011, 2012 and 2015--2018, at centre-of-mass energies of 7, 8 and $13\tev$, respectively. The total data set corresponds to an integrated luminosity of $9\invfb$.
The \lhcb detector is a single-arm forward spectrometer covering the pseudorapidity range $2 < \eta < 5$, designed for the study of particles containing $b$ or $c$ quarks. The detector comprises a high-precision tracking system, including a silicon-strip vertex detector surrounding the $pp$ interaction region, a large-area silicon-strip detector located upstream of a dipole magnet with a bending power of about $4\,\text{T\,m}$, and three stations of silicon-strip detectors and straw drift tubes placed downstream of the magnet. This tracking system provides a measurement of the momentum, $p$, of charged particles with a relative uncertainty ranging from $0.5\%$ at low momentum to $1.0\%$ at $200 \gevc$. The minimum distance of a track to a primary $pp$ collision vertex (PV), known as the impact parameter (IP), is measured with a resolution of $(15 + 29/\pt)\,\mu\text{m}$, where $\pt$ is the component of the momentum transverse to the beam in $\gevc$. Different types of charged hadrons are distinguished using information from two ring-imaging Cherenkov detectors (RICH)~\cite{LHCb-DP-2012-003}.
Photons, electrons and hadrons are identified by a calorimeter system consisting of scintillating-pad and preshower detectors, an electromagnetic calorimeter (ECAL) and a hadronic calorimeter, along with additional information from the RICH.

An online event selection~\cite{LHCb-DP-2012-004, LHCb-DP-2019-001} is performed by a trigger system that consists of a hardware stage, based on information from the calorimeter and muon systems, followed by a software stage, which applies a full event reconstruction.
The hardware electron trigger requires the presence of an ECAL cluster with minimum transverse energy between 2.5 and 3.0 GeV, depending on the data-taking period. Signal \BsPhiEE candidates are retained if at least one of the electrons fulfills the hardware electron trigger criteria or if the trigger requirements are satisfied by objects in the rest of the event that are independent of the signal $\Bs$ candidate decay products. Events meeting the hardware trigger requirements are processed by the software trigger, where a two-, three- or four-track vertex with a significant displacement from any PV is required. At least one charged particle must have a significant transverse momentum and be inconsistent with originating from any PV. A multivariate algorithm is used to identify displaced vertices consistent with the decay of a $b$ hadron~\cite{BBDT}.

Simulated samples are used to assess the effect of reconstruction and selection on the angular distributions of the signal candidates, as well as to estimate expected signal yields and contamination from specific background processes. The $pp$ collisions are simulated using \pythia~\cite{Sjostrand:2007gs} with a specific LHCb configuration~\cite{LHCb-PROC-2010-056}. Decays of unstable particles are described by EvtGen~\cite{Lange:2001uf}, incorporating final-state radiation generated using \photos~\cite{davidson2015photos}. The interaction of the generated particles with the detector, along with its response, is implemented using the \geant toolkit~\cite{Allison:2006ve,LHCb-PROC-2011-006}. Residual mismodelling of the particle identification performance, the $\pt$ spectrum of the $\Bs$ mesons, the track multiplicity, and the efficiency of the hardware trigger are corrected using high-yield control samples from data, following the approach developed by the \lhcb collaboration for tests of lepton universality~\cite{LHCb-PAPER-2022-045}.

\section{Selection of signal candidates}

\subsection{Reconstruction}

The $\phi$ candidates are assembled by combining charged tracks identified as $\Kp$ and $\Km$ mesons, ensuring the invariant mass of the $K^+K^-$ system lies within $\pm$12 $\mevcc$ of the known $\phi$ mass~\cite{PDG2024}. The reconstruction of $\Bs \to \phi \epem$ candidates involves pairing the $\phi$ candidate with a pair of oppositely charged tracks identified as electrons. Due to bremsstrahlung photon emission, electrons lose a substantial amount of energy when interacting with detector material. A dedicated procedure, which searches for neutral transverse energy deposits larger than 75\mev in the ECAL that are compatible with being emitted by an electron upstream of the magnet, is applied to correct for this effect. A comparison of this bremsstrahlung recovery procedure between data and simulated samples in dedicated control channels shows good agreement, thus validating correction method. However, the limitations of this recovery technique degrade the resolution of the reconstructed invariant masses of both the dielectron pair and the $\Bs$ candidate.
The $\Kp\Km\epem$ reconstructed invariant mass is therefore restricted to $4600 < m(\Kp\Km\epem) < 6300 \mevcc$.

The selection of the $\qsq$ region aims to optimise the sensitivity to the photon polarisation in the \btosgam transition present in the \BsPhiEE decay (\ie $C_7^{(')}$). The reconstruction of the invariant mass-squared of the dielectron pair is performed through a kinematic fit~\cite{Hulsbergen:2005pu}, which constrains the $\Kp\Km\epem$ invariant mass to the known $\Bs$ mass~\cite{PDG2024}. This technique, identical to the one developed in Ref.~\cite{LHCb-PAPER-2020-020},
improves the \qsq resolution. Subsequently, the reconstructed \qsq is required not to exceed $0.25\gevgevcccc$ to minimise sensitivity to vector and axial-vector currents ($C_9^{(')}$ and $C_{10}^{(')}$), while maximising the retention of signal candidates. Low-$\qsq$ signal candidates, which are particularly sensitive to $C_7^{(')}$, have a poorer $\Tilde{\varphi}$ resolution due to multiple scattering of quasi-collinear electrons in the tracking detectors. Additionally, contamination from \BsPhiGam decays, followed by photon conversion in the detector material, affects the lower end of the $\qsq$ spectrum. Consequently, a minimum requirement of $0.0001 \gevgevcccc$ is imposed on the reconstructed $\qsq$. The use of the \Bs mass constraint in the kinematic fit to compute the $\qsq$ reduces this background by 80\% with respect to an unconstrained fit. The region of $\qsq$ below $0.0001 \gevgevcccc$, which is enriched in $\Bs \to \phi \gamma$ decays, is used as a control sample and denoted as the ``$\gamma$-bin''.

Each track must exhibit a good quality and be inconsistent with originating from a PV. The kaons must have a transverse momentum larger than $400\mevc$ and be identified as such using information from the RICH detectors. Electrons with \pt exceeding $500\mevc$ and demonstrating a good-quality vertex are utilised to construct dielectron candidates.
In events with multiple PVs, the one giving rise to the smallest value of \chisqip is associated to the \BsPhiEE candidate.

To mitigate background pollution stemming from random track combinations (combinatorial background), a boosted decision tree (BDT) classifier, trained with the XGBoost algorithm~\cite{XGBoost}, is employed. The upper mass sideband of the data samples, defined by \mbox{$m(\Kp\Km\epem) > 5700\mevcc$}, serves as a proxy for the modelling of the combinatorial background, where statistical enrichment is achieved by relaxing the requirement on the invariant mass of the $\Kp\Km$ system (to be between 990 and $1100 \mevcc$). As a signal proxy, simulation samples of $\Bs \to \phi \epem$ candidates are used. A $k$-fold cross-validation approach with $k = 10 $ is used~\cite{10.1145/307400.307439}, in which each of the ten classifiers is trained leaving out a different 10\% of the data sample.
The classifier uses eight kinematic and decay-topology variables, notably including information on the impact parameter of final-state particles, as well as the transverse momentum of the $\Bs$ candidate and its flight distance. Other variables include the angle between the $\Bs$ momentum and its flight distance vector, the fit quality of the $\Bs$ vertex, and its displacement from the associated PV. The classifier reduces the background level by half while retaining 99\% of the signal.

\subsection{Exclusive backgrounds}
Various sources of specific backgrounds are studied using dedicated simulation samples. Identified background sources expected to contribute at a level greater than 1\% of the expected signal are modelled and included in the angular fit described in Sec.~\ref{sec:AngAna}.

A significant background pollution arises from semileptonic \mbox{$\Bs \to \Dsm (\to \phi e^- \Bar{\nu_e}) \ep \nu_e$} decays which share a common final state with the signal given that the neutrinos are not reconstructed. This background populates the region below the $\Bs$ invariant mass and has a combined branching fraction about four orders of magnitude larger than that of the signal~\cite{PDG2024}. Even though the vast majority of this background populates higher \qsq regions, when both neutrinos possess low energies the effectiveness of the signal selection in rejecting this background diminishes. For this background, the positron from the $\Bs$ decay tends to exhibit higher energy than the electron from the $\Dsm$ decay. Consequently, the reconstructed $\cos \theta_L$ distribution tends to favor larger values, as $\cos \theta_L$ is highly correlated with the $\epem$ energy asymmetry. As a result, a symmetric requirement $|\cos \theta_L| < 0.9$ is imposed to further suppress this background. While this leads to a $1\%$ loss of signal, it removes $40\%$ of this semileptonic background, sufficient that it can be subsequently neglected in the fit. Additionally, the contribution from the \mbox{$\Bs \to \Dssm (\to \phi e^- \Bar{\nu_e}) \ep \nu_e$} mode is also studied. It is found to be smaller than the \Dsm one by a factor 0.63, and therefore negligible.

The radiative \BsPhiGam branching fraction is approximately two orders of magnitude larger than that of the signal~\cite{PDG2024}, featuring a very similar distribution in the reconstructed $\Kp\Km\epem$ invariant mass. In the very low \qsq region, the contamination from \BsPhiGam decays is measured to be at the level of $15\%$, but is reduced to about $3\%$ by applying a conversion veto rejecting $\gamma \to \epem$ processes compatible with originating from detector material.

Candidates from \BsPhiEta and \BsPhiPiz decays, where the $\eta$ or $\piz$ meson decays through the $\epem\gamma$ Dalitz process~\cite{dalitz_decay}, can pass the selection if the photon has low energy, or if it is recovered as a bremsstrahlung photon. In the latter case, the reconstructed $\Kp\Km\epem$ invariant-mass peaks at the $\Bs$ invariant mass. The branching fraction  of the $\eta$ and $\piz$ Dalitz decays are assumed to be similar to the corresponding \Bd decays~\cite{PDG2024} and their contamination are estimated to be at the level of $4.7\%$ and $1.8\%$, respectively.

Background contributions from misidentified \mbox{$\Bs\to\phi(\to\Kp\Km)\phi(\to\Kp\Km)$}, \mbox{$\Bd \to \Kstarz(\to\Kp\pi^-) e^+e^-$} and \mbox{$\Lb \to p \Km \epem$} decays are studied but are found to be negligible, due to a low electron misidentification rate ($\sim 2\%$) and the narrowness of the $\phi$ resonance.

\subsection{Effective \boldmath{\qsq} range of the \BsPhiEE signal events}
\begin{figure}
    \centering
    \includegraphics[width=0.8\textwidth]{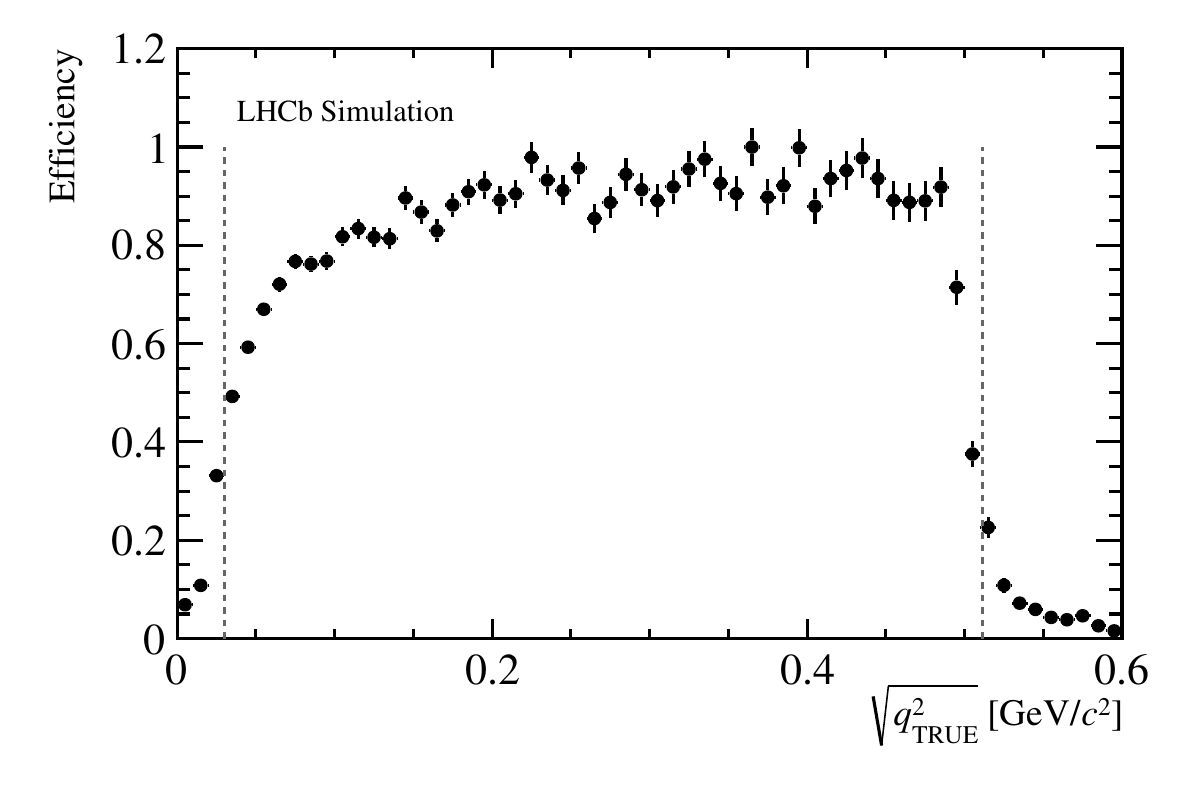}
    \caption{Relative efficiency as a function of the true dielectron invariant mass ($\sqrt{q^2_\text{TRUE}}$). The points represent the efficiency obtained from simulation, while the vertical lines represent the effective boundaries ($\sqrt{q^2_\text{TRUE}}$ between $30\,{\rm MeV\!/}c^2$ and $511\,{\rm MeV\!/}c^2$). The use of $\sqrt{q^2_\text{TRUE}}$ instead of $q^2_{\text{TRUE}}$ allows for better visualisation of the boundary effects.}
    \label{fig:q2eff}
\end{figure}

The dependence of the signal efficiency as a function of the true dielectron invariant mass, as determined through simulation, is shown in Fig.~\ref{fig:q2eff}. This efficiency is slowly varying in most of the signal region.
Close to the boundaries, the efficiency drops rapidly due to the selected range of reconstructed \qsq and the effect of the dielectron mass resolution. Therefore, following the method described in Ref.~\cite{LHCb-PAPER-2014-066}, effective \qsq boundaries are defined to be located between $0.0009 \gevgevcccc$ and $0.2615 \gevgevcccc$ to allow for theoretical predictions of the angular observables without any input from simulation.

\section{Angular analysis \label{sec:AngAna}}
The angular observables \FL, \AT{(2)}, \AT{\Imag \CP} and \AT{\Real \CP} , defined in Eq.~\eqref{eq:PhysicsPDF}, are determined through a four-dimensional unbinned maximum-likelihood fit to the $m(\Kp\Km\epem)$, \ctl, \ctk, and $\Tilde{\varphi}$ distributions.
After the selection, the $\Kp\Km\epem$ invariant mass does not exhibit correlation with the angular distributions as indicated by simulation. The $m(\Kp\Km\epem)$ distribution is thus incorporated into the fit to enhance the discrimination power between signal and background. Both signal and $\Bs\to\phi\gamma$ invariant-mass components are represented by a double-sided Crystal Ball function \cite{Skwarnicki:1986xj} that consists of a Gaussian core with asymmetric power-law tails. The tail parameters are fixed to the values obtained from simulation. The widths and mean values are corrected for differences between data and simulation obtained from the $\gamma$-bin fit results.

The angular distributions of the signal candidates are parameterised by the product of Eq.~\eqref{eq:PhysicsPDF} and a function describing the angular acceptance of the candidate reconstruction and selection process. A dedicated simulated sample generated with flat angular distributions is used to model the angular acceptance, assuming factorization of the three angles: \mbox{$\epsilon(\cos\theta_L, \cos\theta_K, \Tilde{\varphi}) = \epsilon(\cos\theta_L) \times \epsilon(\cos\theta_L) \times \epsilon(\Tilde{\varphi})$}. The acceptance is parameterised using Legendre polynomials of order eight for $\cos\theta_L$ and order four for $\cos\theta_K$.  Given the flavour-symmetric final state, only even orders are considered. For the $\Tilde{\varphi}$ angle, nonuniform acceptance terms proportional to $\cos(2\Tilde{\varphi})$ and
$\sin(2\Tilde{\varphi})$ are allowed. However, no significant deviation from a uniform distribution is observed.

The combinatorial background in $m(\Kp\Km\epem)$ is described by an exponential function whose slope is a free parameter in the fit. The modelling of the angular distribution relies on constraints obtained using
a data sample reconstructed as \mbox{$\Bs \to \phi e^\pm\mu^\mp$}. Indeed, as this decay is forbidden in the SM due to lepton-flavour conservation, the sample mostly comprises semileptonic and combinatorial background candidates. The \qsq and selection requirements are slightly changed to account for the muon in the final state. The $\cos\theta_L$ and $\cos\theta_K$ are modelled separately by Legendre polynomials and the $\Tilde{\varphi}$ distribution is assumed to be uniform. It is verified that the $\Bs \to \phi e^\pm\mu^\mp$ sample is a good proxy for the combinatorial background by comparing the invariant mass and angular distributions to those of \BsPhiEE decays in the lower ($<5100 \mevcc$) and upper ($>5500 \mevcc$) mass sideband regions.

For the \BsPhiGam background, the electrons originate from the interaction of a real photon with detector material. Hence, the \ctl and $\tilde{\varphi}$ dependent parts of Eq.~\eqref{eq:PhysicsPDF} can be integrated out to model only the $\cos\theta_K$ dependent part. This shape is dependent solely on the parameter \FL. The value of \FL for the $\Bs\to\phi\gamma$ decay is obtained from the fit to the $\gamma$-bin. When included as a background in the fit to the signal sample, its angular shape in $\cos\theta_L$ and $\Tilde{\varphi}$ is derived from a simulated sample and is assumed to factorise between the angles.

The shapes of the $\Bs\to\phi\eta$ and $\Bs\to\phi\piz$ background contributions are modelled using simulated samples through nonparametric probability density functions\cite{Cranmer2001} for the invariant mass and via Chebychev polynomials for the angles.

\begin{figure}[b]
    \centering
    \includegraphics[width=0.49\textwidth]{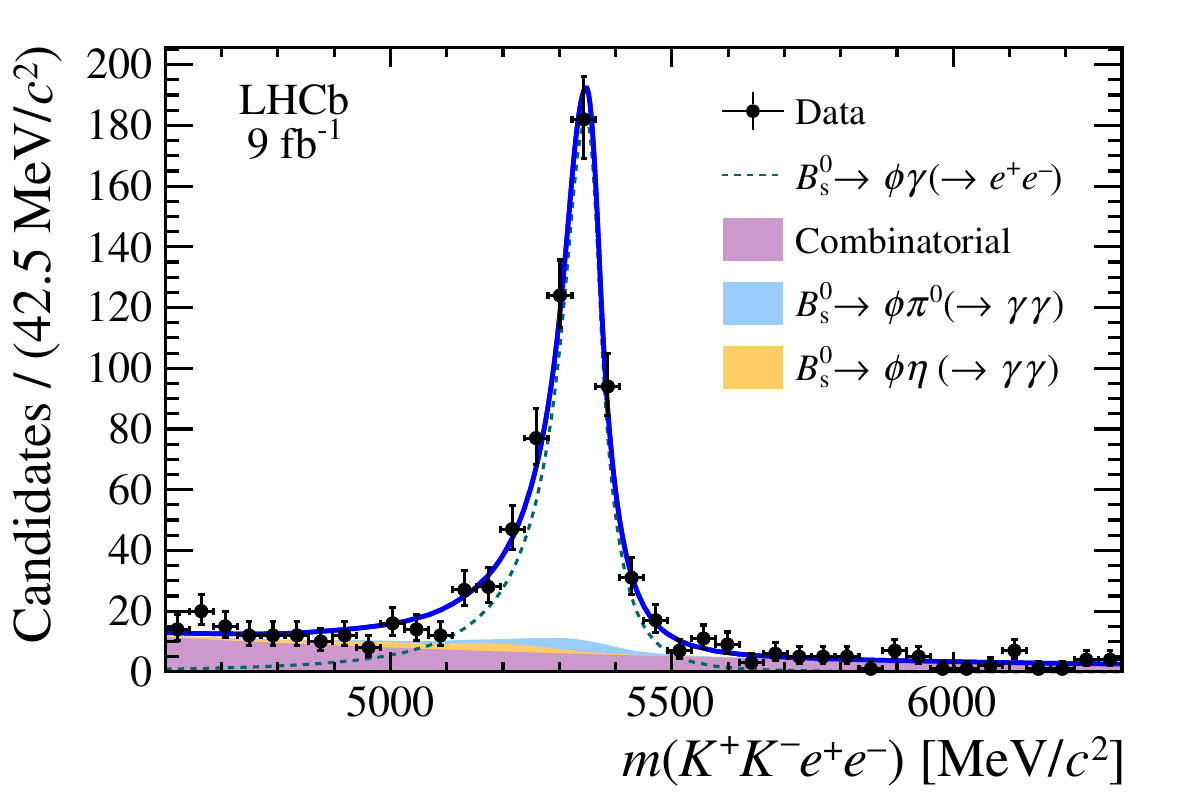}
    \includegraphics[width=0.49\textwidth]{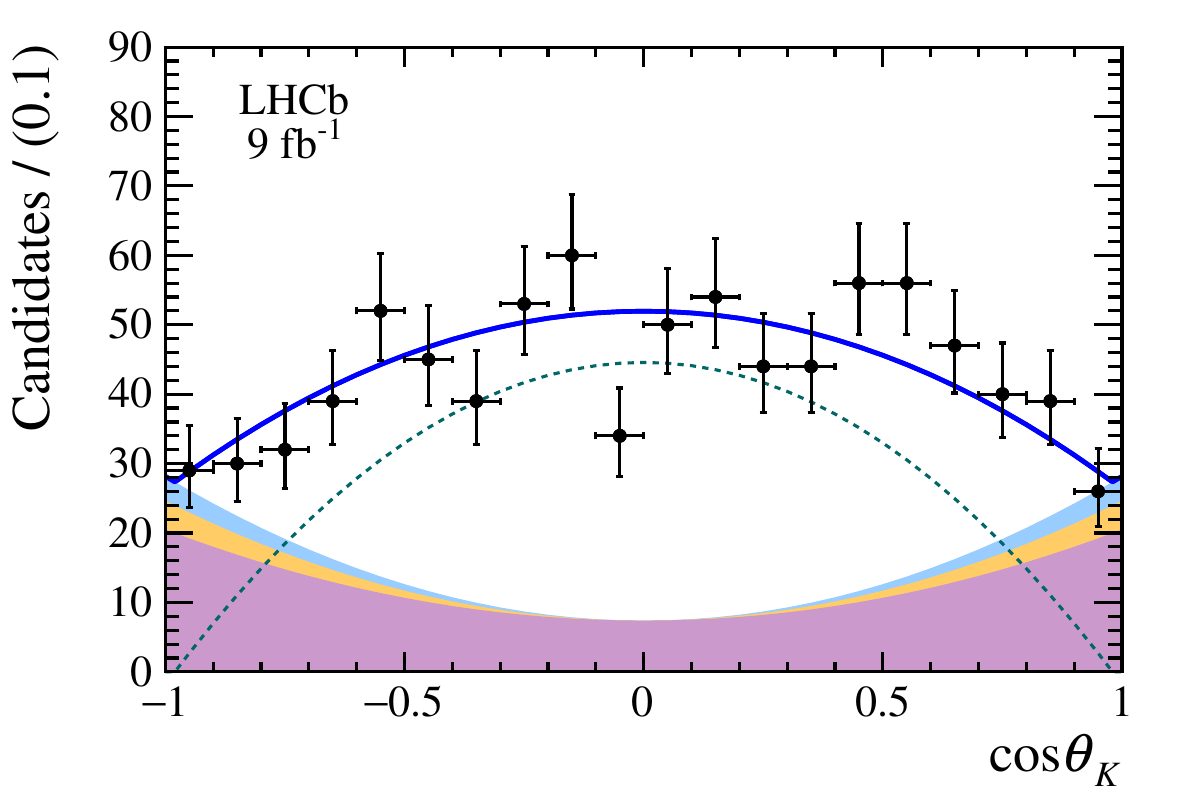}
    \caption{ Distributions of the (left) $K^+K^-e^+e^-$ invariant mass and (right) $\cos{\theta_K}$ of the $B_s^0 \to \phi \gamma$ control sample, with the fit projection also shown.}
    \label{fig:2DFitGammaBin}
\end{figure}

The fitting procedure is validated using a large sample of simulated \BsPhiEE signal decays with a size about 500 times that of the signal yield on data, where the measured values of \FL, \AT{(2)}, \AT{\Imag \CP}, and \AT{\Real \CP} are in agreement with the generated values demonstrating, in particular, the validity of the modelling of the angular acceptance. Subsequently, the \BsPhiEE fit model is validated against data through a two-dimensional fit to the $m(\Kp\Km\epem)$ and \ctk distributions of the \BsPhiGam control sample, corresponding to a total of around $580$ signal candidates. The results from this fit, including their projections, are illustrated in Fig.~\ref{fig:2DFitGammaBin}. The obtained value of $\FL = -0.01 \pm 0.02$, where the uncertainty is statistical only, agrees well with a purely transverse polarisation of the $\phi$ meson, indicative of a real photon interaction.

An additional validation is conducted through pseudoexperiments that include both signal and background components with yields obtained from the data fit, where the angular observables \FL, \AT{(2)} , \AT{\Imag \CP} , and \AT{\Real \CP} are generated according to their SM predictions. The results show good compatibility with the input parameters, with the exception of a minor bias in \FL, quantified as 0.14\% of its statistical uncertainty. This bias is due to the vicinity of the physical boundary at zero, therefore an upper limit on \FL is computed using a Feldman--Cousins approach~\cite{bib:Feldman-Cousins}.

\section{Systematic uncertainties}
A summary of the systematic uncertainties is shown in Table~\ref{tab:Sys}. They are estimated using pseudoexperiments generated according to the baseline result. The pseudoexperiments are fitted with both the default and alternative models, and the resulting difference in angular observables is assigned as the systematic uncertainty.

\begin{table}[tb]
\centering
\begin{tabular}{l|r r r r }
     Source of systematic & \AT{(2)} & \AT{\Imag \CP} & \AT{\Real \CP} & \FL  \\ \hline
    $\Delta\Gamma_s/\Gamma_s$&  0.008& $<$0.001 & $<$0.001 & $<$0.001 \\
    Corrections to simulation    &   0.002&  $<$0.001&   $<$0.001&  0.010 \\
     Acceptance function modelling & $<$0.001& $<$0.001& 0.001& 0.002\\
     Simulation sample size for acceptance & 0.006 &0.008& 0.005& 0.002 \\
    Background contamination  &  0.009& 0.014&  0.004& 0.006\\
    Angles resolution & $-$0.005&  $<$0.001 &  \hfill --- \hfill & \hfill --- \hfill  \\
 \hline
    Total systematic uncertainty & 0.014 & 0.016 & 0.006 & 0.012  \\\hline
    Statistical uncertainty & 0.235 & 0.247 & 0.155 & 0.056 \\ \hline
\end{tabular}
\caption{Summary of the systematic uncertainties. For comparison, the statistical uncertainties are shown in the last row of the table. The dash indicates that the parameter is not affected by the corresponding systematic. }
\label{tab:Sys}
\end{table}
All simulated samples used in the analysis are generated with the $\Delta\Gamma_s$ parameter set to zero. A systematic uncertainty is computed by altering the angular, \qsq, and \Bs lifetime distributions of simulated samples to reflect the current world average value $\frac{\Delta\Gamma_s}{\Gamma_s} = 0.126$~\cite{PDG2024}, primarily affecting the value of \AT{(2)}.

The systematic uncertainties  associated with the corrections applied to simulated candidates used to model the angular acceptance are evaluated by fitting uncorrected simulated candidates. An alternative model using Legendre polynomials with odd coefficients is used to estimate the systematic uncertainties associated with the shape of the acceptance function. To compute the uncertainty linked to the size of the sample used to extract the acceptance shape, the acceptance parameters in pseudoexperiments are randomly generated within one standard deviation.
Peaking background contributions are evaluated by varying the contamination of each background component within its uncertainty.

Furthermore, the non-negligible size of the $\Tilde{\varphi}$ resolution results in an underestimation of the magnitude of \AT{(2)} proportional to its value. The size of the effect is
estimated using signal-only pseudoexperiments generated with a  \qsq-dependent smearing on $\Tilde{\varphi}$ and fitted with the default fitting function which ignores this resolution effect.

It is checked that other sources of systematic uncertainties, such as the signal invariant-mass model and combinatorial angular modelling, yield negligible contributions to the overall systematic uncertainty. This is expected since the related parameters are only loosely constrained in the fit.

The total systematic uncertainties are obtained by summing all contributions in quadrature, and are, in all cases, to be smaller than the statistical uncertainties on the observables.

\section{Results}
\begin{figure}[tb]
    \centering
    \includegraphics[width=0.45\textwidth]{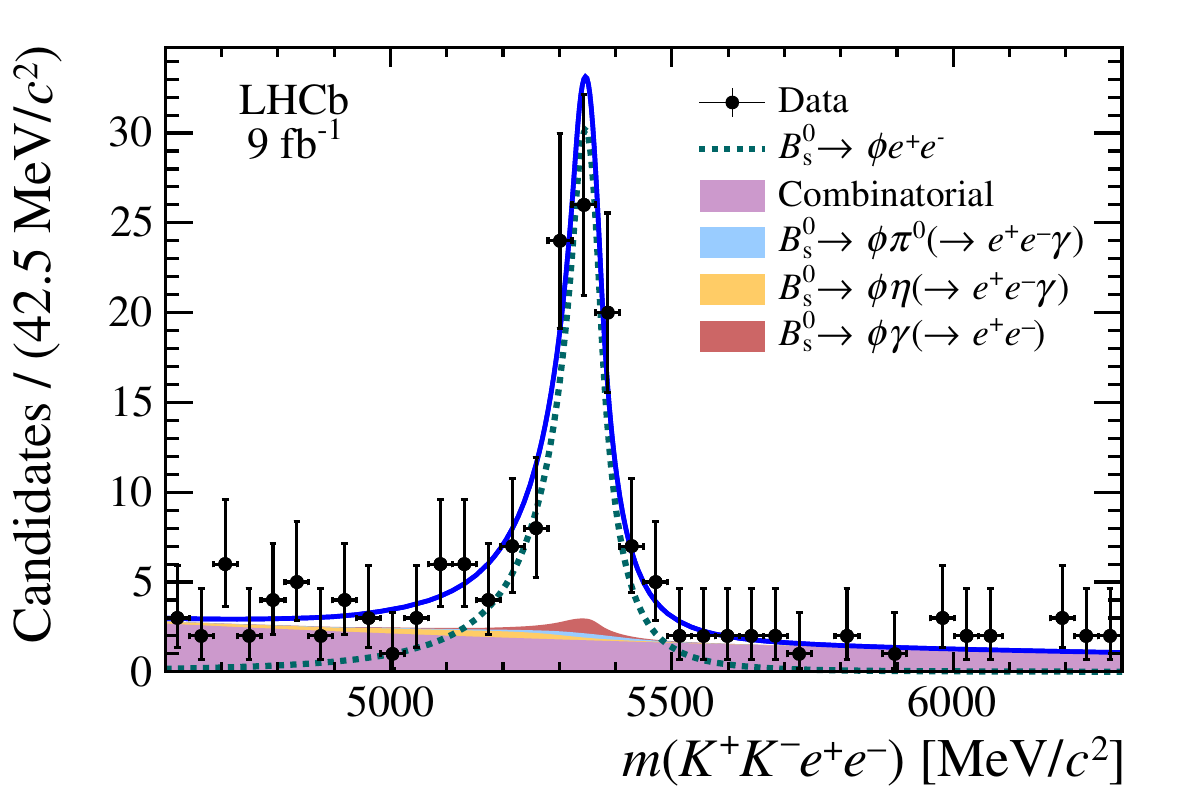}
    \includegraphics[width=0.45\textwidth]{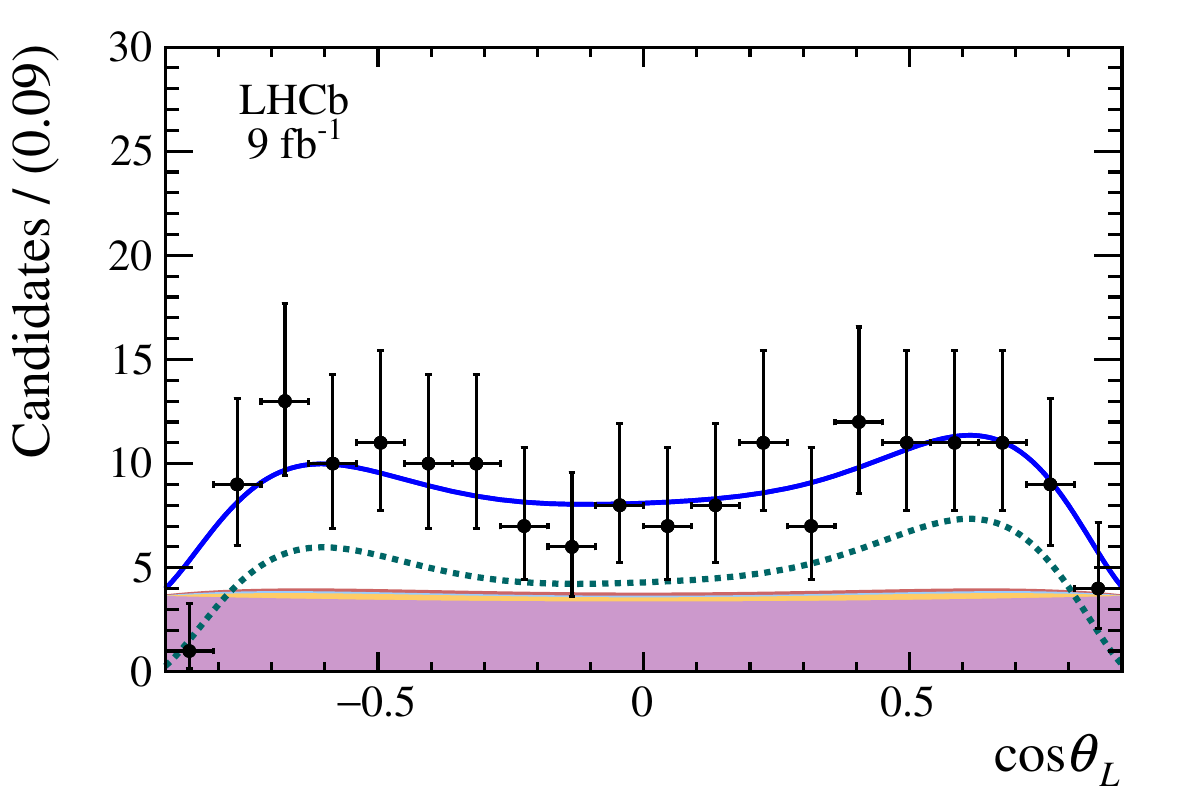}
    \includegraphics[width=0.45\textwidth]{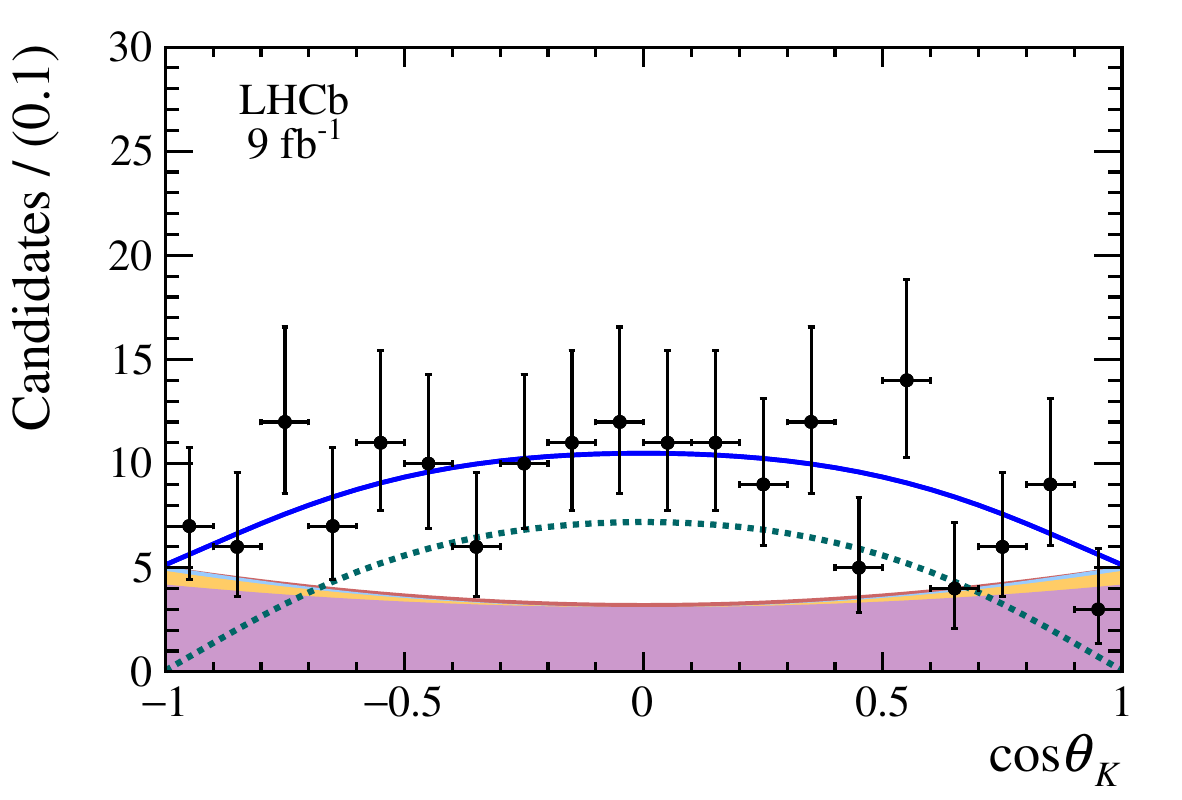}
    \includegraphics[width=0.45\textwidth]{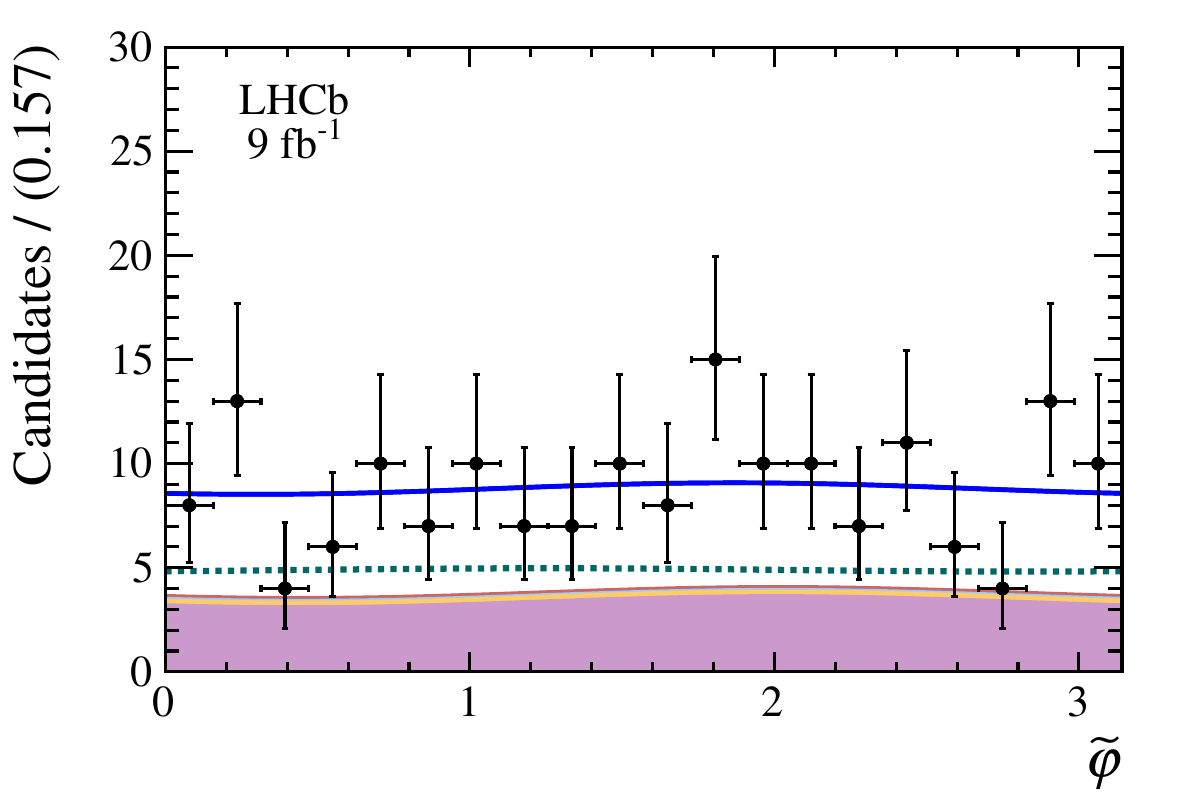}
    \caption{Distributions of the (top left) $K^+K^-e^+e^-$ invariant mass, (top right) $\cos{\theta_L}$, (bottom left) $\cos{\theta_K}$ and (bottom right) $\tilde{\varphi}$ of $B_s^0 \to \phi e^+e^-$ candidates in the very low $q^2$ range with fit projections aslo shown.}
    \label{fig:4DFit}
\end{figure}

The projections of the final fit on $m(\Kp\Km\epem)$ and the three angles are displayed in Fig.~\ref{fig:4DFit}. The total signal yield observed within the effective \qsq range, from $0.0009$ to $0.2615 \gevgevcccc$, is about 100 events. Together with the results from Ref.~\cite{LHCb-PAPER-2024-032}, this constitutes the first observation of the \BsPhiEE decay mode.

\begin{figure}[tb]
  \begin{center}
    \includegraphics[ width=0.7\textwidth]{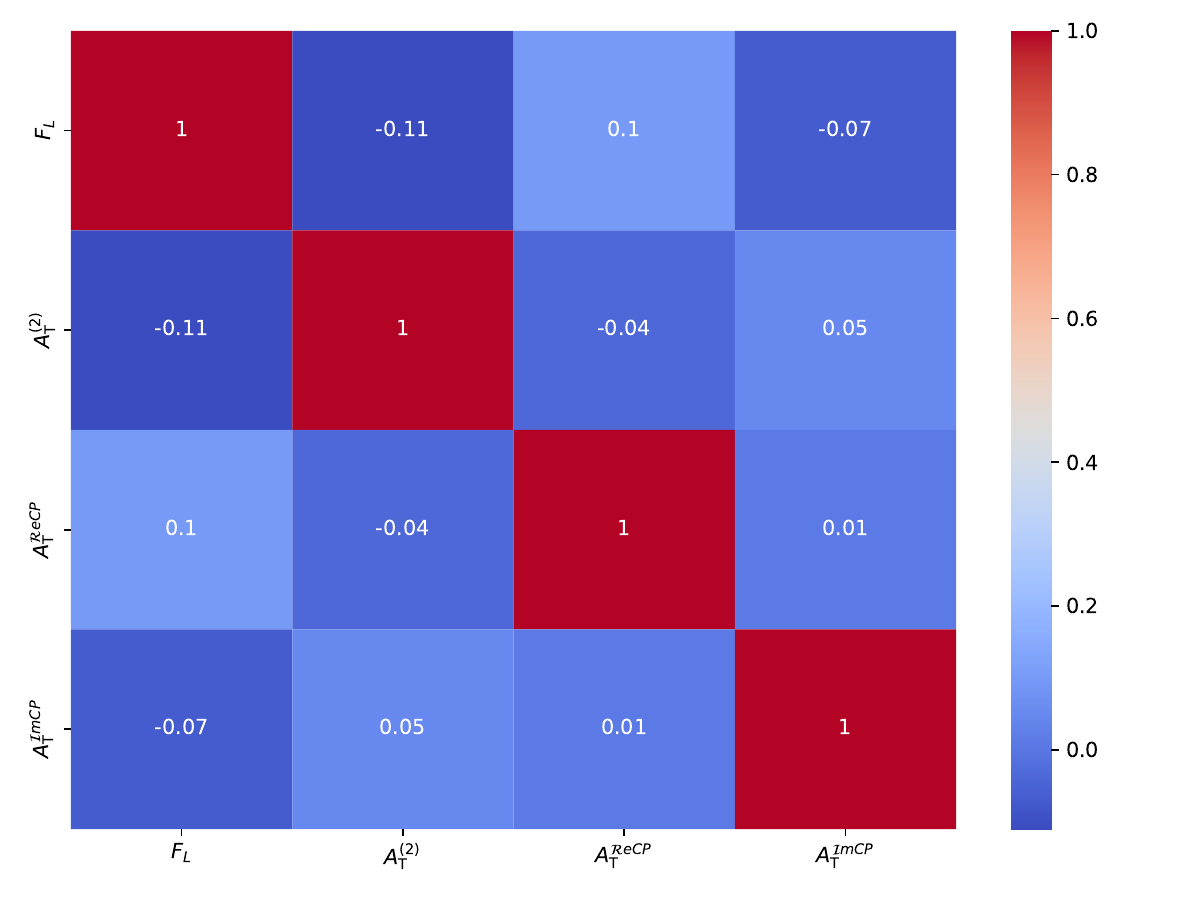}
  \end{center}
  \caption{
    \small
    The statistical correlation matrix for the four angular parameters in the final fit. }
  \label{fig:Corr}
\end{figure}

The obtained values for the four angular observables are
\begin{equation*}
    \begin{aligned}
        \AT^{(2)} &= -0.045 \pm 0.235 \pm 0.014\, , \\
        \AT{\Imag \CP} &= \phantom{+} 0.002 \pm 0.247 \pm 0.016\, , \\
        \AT{\Real \CP} &= \phantom{+} 0.116 \pm 0.155 \pm 0.006\, , \\
        \FL &= \phantom{+} (0.4 \pm 5.6 \pm 1.2) \% \, , \\
    \end{aligned}
\end{equation*}
where the first uncertainty is statistical and the second systematic. Fig.~\ref{fig:Corr} presents the heatmap of the correlation matrix for the four angular parameters from the final fit. Given the full dominance of the statistical uncertainties, it is not significantly altered by the systematical ones. As the \FL parameter is close to the physical boundary, the Feldman--Cousins method is applied to determine an upper limit, $\FL < 11.5\%~(13.7\%) \text{ at the 90\%~(95\%) confidence level (CL)}$.
A shift of $-0.025$ has been applied to the  \AT{(2)} parameter to correct for the remaining bias observed while fitting the simulation corrected for the $\Delta \Gamma_s$ being nonzero. Studies with simulation show that this bias arises from the nonuniform efficiency in the \Bs decay time.

\section{Summary}
\begin{figure}[tb]
    \centering    \includegraphics{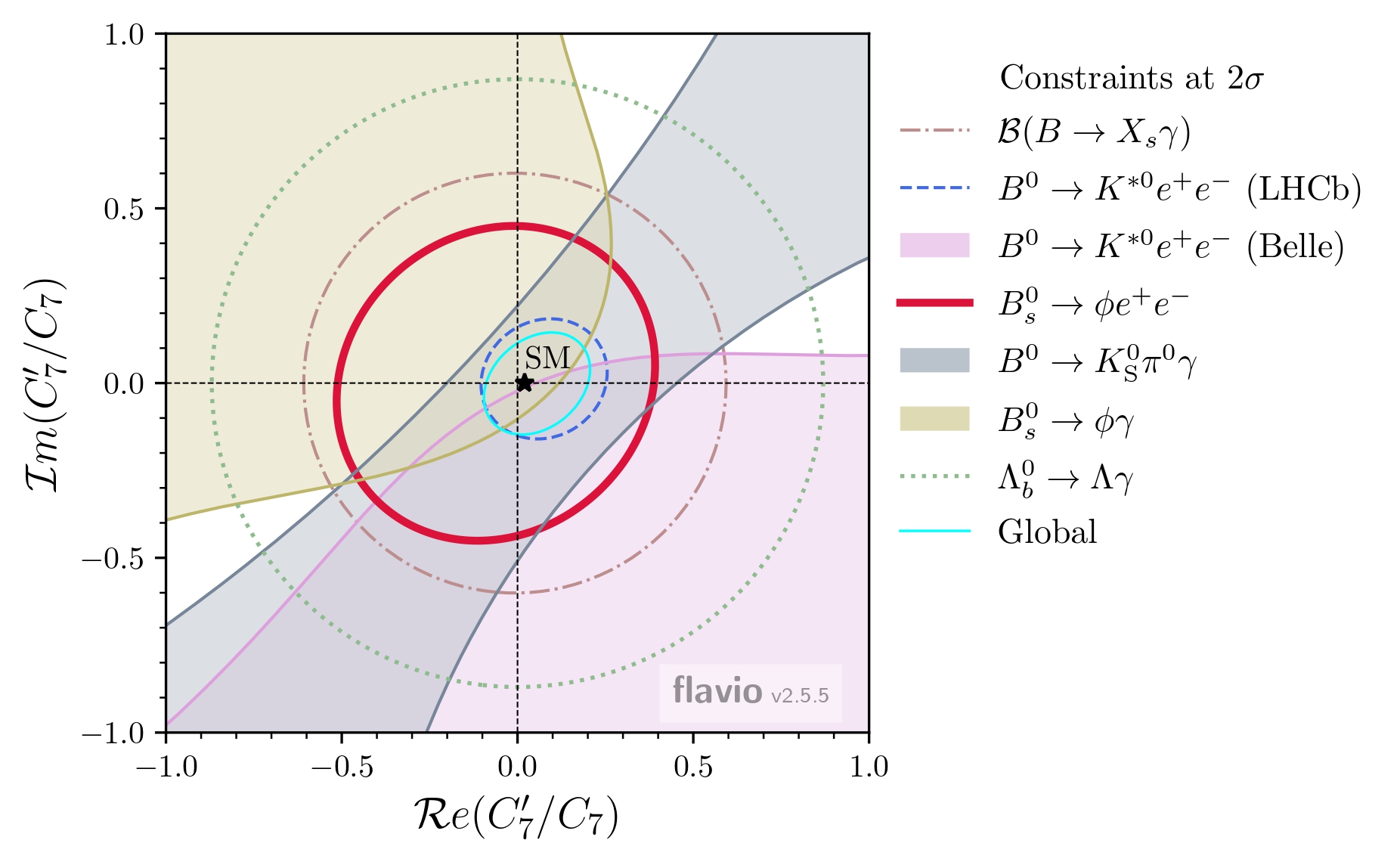}
    \caption{Current constraints at the $2\sigma$ level (containing 95.4\% of the distribution) on the real and imaginary part of the ratio of the right- to left-handed Wilson coefficients $C_7^{'}$ and $C_7$. The constraints from various measurements are shown in light colours and are, for most of them, the combination of several results.  The constraints from angular analyses of $B^0 \to K^{*0}e^+e^-$ decays from the LHCb and the Belle experiments are performed in different $q^2$ regions and are shown separately. The constraints from the measurement presented in this paper are shown in red and the result of the global fit in cyan. The  SM prediction is represented by the black star.}
        \label{fig:FlavioConstraints}
\end{figure}

An angular analysis of the \BsPhiEE decay is performed for the first time, using $pp$ collision data collected by the LHCb experiment between 2011 and 2018, corresponding to an integrated luminosity of 9 \invfb. Angular observables are measured in an effective \qsq region between $0.0009$ and $0.2615\,\gevgevcccc$. The results are consistent with the SM predictions~\cite{Be_irevi__2012,straub2018flavio,Jager:2014rwa,Descotes-Genon:2022qce}. Using the \texttt{flavio} software package~\cite{straub2018flavio}, they can be used to measure both the real and imaginary parts of the \BsPhiGam photon polarisation with a precision of 12\%.
The $C_7^{(')}$ regularisation-scheme-independent effective coefficients are calculated at the scale $\mu = 4.8 \gev$~\cite{Paul2017} and the value of the left-handed $C_7$ coefficient is fixed to its SM value, $-0.2915$. The overall constraints, shown in
Fig.~\ref{fig:FlavioConstraints}, are compared to those from previous measurements from the \babar, Belle, Belle II and LHCb collaborations \cite{Aubert_2008,Lees_2012, Lees_2012_2,Saito_2015,bellecollaboration2016measurement,
Ushiroda_2006, LHCb-PAPER-2019-015, LHCb-PAPER-2021-030, Belle:2024mml, LHCb-PAPER-2020-020,Belle-II:2024uzp}.

\newpage

\newpage
\section*{Acknowledgements}
%
%
\noindent We express our gratitude to our colleagues in the CERN
accelerator departments for the excellent performance of the LHC. We
thank the technical and administrative staff at the LHCb
institutes.
We acknowledge support from CERN and from the national agencies:
CAPES, CNPq, FAPERJ and FINEP (Brazil); 
MOST and NSFC (China); 
CNRS/IN2P3 (France); 
BMBF, DFG and MPG (Germany); 
INFN (Italy); 
NWO (Netherlands); 
MNiSW and NCN (Poland); 
MCID/IFA (Romania); 
MICIU and AEI (Spain);
SNSF and SER (Switzerland); 
NASU (Ukraine); 
STFC (United Kingdom); 
DOE NP and NSF (USA).
We acknowledge the computing resources that are provided by CERN, IN2P3
(France), KIT and DESY (Germany), INFN (Italy), SURF (Netherlands),
PIC (Spain), GridPP (United Kingdom), 
CSCS (Switzerland), IFIN-HH (Romania), CBPF (Brazil),
and Polish WLCG (Poland).
We are indebted to the communities behind the multiple open-source
software packages on which we depend.
Individual groups or members have received support from
ARC and ARDC (Australia);
Key Research Program of Frontier Sciences of CAS, CAS PIFI, CAS CCEPP, 
Fundamental Research Funds for the Central Universities, 
and Sci. \& Tech. Program of Guangzhou (China);
Minciencias (Colombia);
EPLANET, Marie Sk\l{}odowska-Curie Actions, ERC and NextGenerationEU (European Union);
A*MIDEX, ANR, IPhU and Labex P2IO, and R\'{e}gion Auvergne-Rh\^{o}ne-Alpes (France);
AvH Foundation (Germany);
ICSC (Italy); 
Severo Ochoa and Mar\'ia de Maeztu Units of Excellence, GVA, XuntaGal, GENCAT, InTalent-Inditex and Prog. ~Atracci\'on Talento CM (Spain);
SRC (Sweden);
the Leverhulme Trust, the Royal Society
 and UKRI (United Kingdom).


\newpage
\addcontentsline{toc}{section}{References}
\bibliographystyle{LHCb}
\bibliography{main,standard,LHCb-PAPER,LHCb-CONF,LHCb-DP,LHCb-TDR}

\ifx\mcitethebibliography\mciteundefinedmacro
\PackageError{LHCb.bst}{mciteplus.sty has not been loaded}
{This bibstyle requires the use of the mciteplus package.}\fi
\providecommand{\href}[2]{#2}
\begin{mcitethebibliography}{10}
\mciteSetBstSublistMode{n}
\mciteSetBstMaxWidthForm{subitem}{\alph{mcitesubitemcount})}
\mciteSetBstSublistLabelBeginEnd{\mcitemaxwidthsubitemform\space}
{\relax}{\relax}

\bibitem{Atwood:1997zr}
D.~Atwood, M.~Gronau, and A.~Soni,
  \ifthenelse{\boolean{articletitles}}{\emph{{Mixing-induced CP asymmetries in
  radiative B decays in and beyond the standard model}},
  }{}\href{https://doi.org/10.1103/PhysRevLett.79.185}{Phys.\ Rev.\ Lett.\
  \textbf{79} (1997) 185},
  \href{http://arxiv.org/abs/hep-ph/9704272}{{\normalfont\ttfamily
  arXiv:hep-ph/9704272}}\relax
\mciteBstWouldAddEndPuncttrue
\mciteSetBstMidEndSepPunct{\mcitedefaultmidpunct}
{\mcitedefaultendpunct}{\mcitedefaultseppunct}\relax
\EndOfBibitem
\bibitem{Everett:2001yy}
L.~L. Everett {\em et~al.},
  \ifthenelse{\boolean{articletitles}}{\emph{{Alternative approach to $b\to
  s\gamma$ in the uMSSM}},
  }{}\href{https://doi.org/10.1088/1126-6708/2002/01/022}{JHEP \textbf{01}
  (2002) 022}, \href{http://arxiv.org/abs/hep-ph/0112126}{{\normalfont\ttfamily
  arXiv:hep-ph/0112126}}\relax
\mciteBstWouldAddEndPuncttrue
\mciteSetBstMidEndSepPunct{\mcitedefaultmidpunct}
{\mcitedefaultendpunct}{\mcitedefaultseppunct}\relax
\EndOfBibitem
\bibitem{Grinstein:2004uu}
B.~Grinstein, Y.~Grossman, Z.~Ligeti, and D.~Pirjol,
  \ifthenelse{\boolean{articletitles}}{\emph{{Photon polarization in $B \to X
  \gamma$ in the standard model}},
  }{}\href{https://doi.org/10.1103/PhysRevD.71.011504}{Phys.\ Rev.\
  \textbf{D71} (2005) 011504},
  \href{http://arxiv.org/abs/hep-ph/0412019}{{\normalfont\ttfamily
  arXiv:hep-ph/0412019}}\relax
\mciteBstWouldAddEndPuncttrue
\mciteSetBstMidEndSepPunct{\mcitedefaultmidpunct}
{\mcitedefaultendpunct}{\mcitedefaultseppunct}\relax
\EndOfBibitem
\bibitem{Foster:2006ze}
J.~Foster, K.-i. Okumura, and L.~Roszkowski,
  \ifthenelse{\boolean{articletitles}}{\emph{{New constraints on SUSY flavour
  mixing in light of recent measurements at the Tevatron}},
  }{}\href{https://doi.org/10.1016/j.physletb.2006.09.004}{Phys.\ Lett.\
  \textbf{B641} (2006) 452},
  \href{http://arxiv.org/abs/hep-ph/0604121}{{\normalfont\ttfamily
  arXiv:hep-ph/0604121}}\relax
\mciteBstWouldAddEndPuncttrue
\mciteSetBstMidEndSepPunct{\mcitedefaultmidpunct}
{\mcitedefaultendpunct}{\mcitedefaultseppunct}\relax
\EndOfBibitem
\bibitem{Lunghi:2006hc}
E.~Lunghi and J.~Matias, \ifthenelse{\boolean{articletitles}}{\emph{{Huge
  right-handed current effects in $\B\to\Kstarz(K\pi^0)\ell^+\ell^-$ in
  supersymmetry}}, }{}\href{https://doi.org/10.1088/1126-6708/2007/04/058}{JHEP
  \textbf{04} (2007) 058},
  \href{http://arxiv.org/abs/hep-ph/0612166}{{\normalfont\ttfamily
  arXiv:hep-ph/0612166}}\relax
\mciteBstWouldAddEndPuncttrue
\mciteSetBstMidEndSepPunct{\mcitedefaultmidpunct}
{\mcitedefaultendpunct}{\mcitedefaultseppunct}\relax
\EndOfBibitem
\bibitem{Becirevic:2012dx}
D.~Bečirević, E.~Kou, A.~Le~Yaouanc, and A.~Tayduganov,
  \ifthenelse{\boolean{articletitles}}{\emph{{Future prospects for the
  determination of the Wilson coefficient $C_{7\gamma}^\prime$}},
  }{}\href{https://doi.org/10.1007/JHEP08(2012)090}{JHEP \textbf{08} (2012)
  090}, \href{http://arxiv.org/abs/1206.1502}{{\normalfont\ttfamily
  arXiv:1206.1502}}\relax
\mciteBstWouldAddEndPuncttrue
\mciteSetBstMidEndSepPunct{\mcitedefaultmidpunct}
{\mcitedefaultendpunct}{\mcitedefaultseppunct}\relax
\EndOfBibitem
\bibitem{Aubert_2008}
\babar collaboration, B.~Aubert {\em et~al.},
  \ifthenelse{\boolean{articletitles}}{\emph{{Measurement of the
  \decay{B}{X_s\gamma} branching fraction and photon energy spectrum using the
  recoil method}}, }{}\href{https://doi.org/10.1103/physrevd.77.051103}{Phys.\
  Rev.\  \textbf{D77} (2008) 051103},
  \href{http://arxiv.org/abs/0711.4889}{{\normalfont\ttfamily
  arXiv:0711.4889}}\relax
\mciteBstWouldAddEndPuncttrue
\mciteSetBstMidEndSepPunct{\mcitedefaultmidpunct}
{\mcitedefaultendpunct}{\mcitedefaultseppunct}\relax
\EndOfBibitem
\bibitem{Lees_2012}
\babar collaboration, J.~P. Lees {\em et~al.},
  \ifthenelse{\boolean{articletitles}}{\emph{{Precision measurement of the
  \decay{B}{X_s\gamma} photon energy spectrum, branching fraction and direct
  \CP asymmetry A$_{CP}$(\decay{B}{X_{s+d}\gamma})}},
  }{}\href{https://doi.org/10.1103/physrevlett.109.191801}{Phys.\ Rev.\ Lett.\
  \textbf{109} (2012) 191801},
  \href{http://arxiv.org/abs/1207.2690}{{\normalfont\ttfamily
  arXiv:1207.2690}}\relax
\mciteBstWouldAddEndPuncttrue
\mciteSetBstMidEndSepPunct{\mcitedefaultmidpunct}
{\mcitedefaultendpunct}{\mcitedefaultseppunct}\relax
\EndOfBibitem
\bibitem{Lees_2012_2}
\babar collaboration, J.~P. Lees {\em et~al.},
  \ifthenelse{\boolean{articletitles}}{\emph{{Exclusive measurements of
  \btosgam transition rate and photon energy spectrum}},
  }{}\href{https://doi.org/10.1103/physrevd.86.052012}{Phys.\ Rev.\
  \textbf{D86} (2012) 052012},
  \href{http://arxiv.org/abs/1207.2520}{{\normalfont\ttfamily
  arXiv:1207.2520}}\relax
\mciteBstWouldAddEndPuncttrue
\mciteSetBstMidEndSepPunct{\mcitedefaultmidpunct}
{\mcitedefaultendpunct}{\mcitedefaultseppunct}\relax
\EndOfBibitem
\bibitem{Saito_2015}
Belle collaboration, T.~Saito {\em et~al.},
  \ifthenelse{\boolean{articletitles}}{\emph{{Measurement of the
  \decay{\Bbar}{X_s\gamma} branching fraction with a sum of exclusive decays}},
  }{}\href{https://doi.org/10.1103/physrevd.91.052004}{Phys.\ Rev.\
  \textbf{D91} (2015) 052004},
  \href{http://arxiv.org/abs/1411.7198}{{\normalfont\ttfamily
  arXiv:1411.7198}}\relax
\mciteBstWouldAddEndPuncttrue
\mciteSetBstMidEndSepPunct{\mcitedefaultmidpunct}
{\mcitedefaultendpunct}{\mcitedefaultseppunct}\relax
\EndOfBibitem
\bibitem{bellecollaboration2016measurement}
Belle collaboration, \ifthenelse{\boolean{articletitles}}{\emph{{Measurement of
  the inclusive \decay{\B}{X_{s+d}\gamma} branching fraction, photon energy
  spectrum and HQE parameters}},
  }{}\href{http://arxiv.org/abs/1608.02344}{{\normalfont\ttfamily
  arXiv:1608.02344}}\relax
\mciteBstWouldAddEndPuncttrue
\mciteSetBstMidEndSepPunct{\mcitedefaultmidpunct}
{\mcitedefaultendpunct}{\mcitedefaultseppunct}\relax
\EndOfBibitem
\bibitem{Ushiroda_2006}
Belle collaboration, Y.~Ushiroda {\em et~al.},
  \ifthenelse{\boolean{articletitles}}{\emph{{Time-dependent CP asymmetries in
  \decay{\B}{\KS\piz\gamma} transitions}},
  }{}\href{https://doi.org/10.1103/physrevd.74.111104}{Phys.\ Rev.\
  \textbf{D74} (2006) 111104},
  \href{http://arxiv.org/abs/hep-ex/0608017}{{\normalfont\ttfamily
  arXiv:hep-ex/0608017}}\relax
\mciteBstWouldAddEndPuncttrue
\mciteSetBstMidEndSepPunct{\mcitedefaultmidpunct}
{\mcitedefaultendpunct}{\mcitedefaultseppunct}\relax
\EndOfBibitem
\bibitem{Belle:2024mml}
Belle Collaboration, D.~Ferlewicz {\em et~al.},
  \ifthenelse{\boolean{articletitles}}{\emph{{Angular analysis of
  $B\ensuremath{\rightarrow}{K}^{*}{e}^{+}{e}^{\ensuremath{-}}$ in the
  low-${q}^{2}$ region with new electron identification at Belle}},
  }{}\href{https://doi.org/10.1103/PhysRevD.110.072005}{Phys.\ Rev.\ D
  \textbf{110} (2024) 072005},
  \href{http://arxiv.org/abs/2404.00201}{{\normalfont\ttfamily
  arXiv:2404.00201}}\relax
\mciteBstWouldAddEndPuncttrue
\mciteSetBstMidEndSepPunct{\mcitedefaultmidpunct}
{\mcitedefaultendpunct}{\mcitedefaultseppunct}\relax
\EndOfBibitem
\bibitem{Belle-II:2024uzp}
Belle II collaboration, I.~Adachi {\em et~al.},
  \ifthenelse{\boolean{articletitles}}{\emph{{Measurement of $CP$ asymmetries
  in $B^0 \to K^0_S \pi^0 \gamma$ decays at Belle II}},
  }{}\href{http://arxiv.org/abs/2407.09139}{{\normalfont\ttfamily
  arXiv:2407.09139}}\relax
\mciteBstWouldAddEndPuncttrue
\mciteSetBstMidEndSepPunct{\mcitedefaultmidpunct}
{\mcitedefaultendpunct}{\mcitedefaultseppunct}\relax
\EndOfBibitem
\bibitem{LHCb-PAPER-2019-015}
LHCb collaboration, R.~Aaij {\em et~al.},
  \ifthenelse{\boolean{articletitles}}{\emph{{Measurement of \CP-violating and
  mixing-induced observables in \mbox{\decay{\Bs}{\phi \gamma}} decays}},
  }{}\href{https://doi.org/10.1103/PhysRevLett.123.081802}{Phys.\ Rev.\ Lett.\
  \textbf{123} (2019) 081802},
  \href{http://arxiv.org/abs/1905.06284}{{\normalfont\ttfamily
  arXiv:1905.06284}}\relax
\mciteBstWouldAddEndPuncttrue
\mciteSetBstMidEndSepPunct{\mcitedefaultmidpunct}
{\mcitedefaultendpunct}{\mcitedefaultseppunct}\relax
\EndOfBibitem
\bibitem{LHCb-PAPER-2021-030}
LHCb collaboration, R.~Aaij {\em et~al.},
  \ifthenelse{\boolean{articletitles}}{\emph{{Measurement of the photon
  polarization in $\Lb \to \Lz \gamma$ decays}},
  }{}\href{https://doi.org/10.1103/PhysRevD.105.L051104}{Phys.\ Rev.\
  \textbf{D105} (2022) 051104},
  \href{http://arxiv.org/abs/2111.10194}{{\normalfont\ttfamily
  arXiv:2111.10194}}\relax
\mciteBstWouldAddEndPuncttrue
\mciteSetBstMidEndSepPunct{\mcitedefaultmidpunct}
{\mcitedefaultendpunct}{\mcitedefaultseppunct}\relax
\EndOfBibitem
\bibitem{LHCb-PAPER-2020-020}
LHCb collaboration, R.~Aaij {\em et~al.},
  \ifthenelse{\boolean{articletitles}}{\emph{{Strong constraints on the $b \to
  s\gamma$ photon polarisation from $B^0 \to K^{\ast 0} e^+ e^-$ decays}},
  }{}\href{https://doi.org/10.1007/JHEP12(2020)081}{JHEP \textbf{12} (2020)
  081}, \href{http://arxiv.org/abs/2010.06011}{{\normalfont\ttfamily
  arXiv:2010.06011}}\relax
\mciteBstWouldAddEndPuncttrue
\mciteSetBstMidEndSepPunct{\mcitedefaultmidpunct}
{\mcitedefaultendpunct}{\mcitedefaultseppunct}\relax
\EndOfBibitem
\bibitem{Bharucha_2016}
A.~Bharucha, D.~M. Straub, and R.~Zwicky,
  \ifthenelse{\boolean{articletitles}}{\emph{{\decay{B}{V\ell^+\ell^-} in the
  Standard Model from light-cone sum rules}},
  }{}\href{https://doi.org/10.1007/jhep08(2016)098}{JHEP \textbf{2016} (2016)
  098}, \href{http://arxiv.org/abs/1503.05534}{{\normalfont\ttfamily
  arXiv:1503.05534}}\relax
\mciteBstWouldAddEndPuncttrue
\mciteSetBstMidEndSepPunct{\mcitedefaultmidpunct}
{\mcitedefaultendpunct}{\mcitedefaultseppunct}\relax
\EndOfBibitem
\bibitem{Ball2007}
P.~Ball, G.~W. Jones, and R.~Zwicky,
  \ifthenelse{\boolean{articletitles}}{\emph{{\decay{B}{V \gamma} beyond QCD
  factorization}}, }{}\href{https://doi.org/10.1103/physrevd.75.054004}{Phys.\
  Rev.\  \textbf{D75} (2007) },
  \href{http://arxiv.org/abs/hep-ph/0612081}{{\normalfont\ttfamily
  arXiv:hep-ph/0612081}}\relax
\mciteBstWouldAddEndPuncttrue
\mciteSetBstMidEndSepPunct{\mcitedefaultmidpunct}
{\mcitedefaultendpunct}{\mcitedefaultseppunct}\relax
\EndOfBibitem
\bibitem{LHCb-PAPER-2014-066}
LHCb collaboration, R.~Aaij {\em et~al.},
  \ifthenelse{\boolean{articletitles}}{\emph{{Angular analysis of the
  \mbox{\decay{\Bz}{\Kstar\epem}} decay in the low-$q^2$ region}},
  }{}\href{https://doi.org/10.1007/JHEP04(2015)064}{JHEP \textbf{04} (2015)
  064}, \href{http://arxiv.org/abs/1501.03038}{{\normalfont\ttfamily
  arXiv:1501.03038}}\relax
\mciteBstWouldAddEndPuncttrue
\mciteSetBstMidEndSepPunct{\mcitedefaultmidpunct}
{\mcitedefaultendpunct}{\mcitedefaultseppunct}\relax
\EndOfBibitem
\bibitem{Descotes-Genon:2022qce}
S.~Descotes-Genon, I.~Plakias, and O.~Sumensari,
  \ifthenelse{\boolean{articletitles}}{\emph{{On the impact of meson mixing on
  B$_{s}$\textrightarrow{} \ensuremath{\phi}ee angular observables at low
  q$^{2}$}}, }{}\href{https://doi.org/10.1007/JHEP02(2023)096}{JHEP \textbf{02}
  (2023) 096}, \href{http://arxiv.org/abs/2210.11995}{{\normalfont\ttfamily
  arXiv:2210.11995}}\relax
\mciteBstWouldAddEndPuncttrue
\mciteSetBstMidEndSepPunct{\mcitedefaultmidpunct}
{\mcitedefaultendpunct}{\mcitedefaultseppunct}\relax
\EndOfBibitem
\bibitem{LHCb-DP-2012-003}
M.~Adinolfi {\em et~al.},
  \ifthenelse{\boolean{articletitles}}{\emph{{Performance of the \lhcb RICH
  detector at the LHC}},
  }{}\href{https://doi.org/10.1140/epjc/s10052-013-2431-9}{Eur.\ Phys.\ J.\
  \textbf{C73} (2013) 2431},
  \href{http://arxiv.org/abs/1211.6759}{{\normalfont\ttfamily
  arXiv:1211.6759}}\relax
\mciteBstWouldAddEndPuncttrue
\mciteSetBstMidEndSepPunct{\mcitedefaultmidpunct}
{\mcitedefaultendpunct}{\mcitedefaultseppunct}\relax
\EndOfBibitem
\bibitem{LHCb-DP-2012-004}
R.~Aaij {\em et~al.}, \ifthenelse{\boolean{articletitles}}{\emph{{The \lhcb
  trigger and its performance in 2011}},
  }{}\href{https://doi.org/10.1088/1748-0221/8/04/P04022}{JINST \textbf{8}
  (2013) P04022}, \href{http://arxiv.org/abs/1211.3055}{{\normalfont\ttfamily
  arXiv:1211.3055}}\relax
\mciteBstWouldAddEndPuncttrue
\mciteSetBstMidEndSepPunct{\mcitedefaultmidpunct}
{\mcitedefaultendpunct}{\mcitedefaultseppunct}\relax
\EndOfBibitem
\bibitem{LHCb-DP-2019-001}
R.~Aaij {\em et~al.}, \ifthenelse{\boolean{articletitles}}{\emph{{Design and
  performance of the LHCb trigger and full real-time reconstruction in Run 2 of
  the LHC}}, }{}\href{https://doi.org/10.1088/1748-0221/14/04/P04013}{JINST
  \textbf{14} (2019) P04013},
  \href{http://arxiv.org/abs/1812.10790}{{\normalfont\ttfamily
  arXiv:1812.10790}}\relax
\mciteBstWouldAddEndPuncttrue
\mciteSetBstMidEndSepPunct{\mcitedefaultmidpunct}
{\mcitedefaultendpunct}{\mcitedefaultseppunct}\relax
\EndOfBibitem
\bibitem{BBDT}
V.~V. Gligorov and M.~Williams,
  \ifthenelse{\boolean{articletitles}}{\emph{{Efficient, reliable and fast
  high-level triggering using a bonsai boosted decision tree}},
  }{}\href{https://doi.org/10.1088/1748-0221/8/02/P02013}{JINST \textbf{8}
  (2013) P02013}, \href{http://arxiv.org/abs/1210.6861}{{\normalfont\ttfamily
  arXiv:1210.6861}}\relax
\mciteBstWouldAddEndPuncttrue
\mciteSetBstMidEndSepPunct{\mcitedefaultmidpunct}
{\mcitedefaultendpunct}{\mcitedefaultseppunct}\relax
\EndOfBibitem
\bibitem{Sjostrand:2007gs}
T.~Sj\"{o}strand, S.~Mrenna, and P.~Skands,
  \ifthenelse{\boolean{articletitles}}{\emph{{A brief introduction to PYTHIA
  8.1}}, }{}\href{https://doi.org/10.1016/j.cpc.2008.01.036}{Comput.\ Phys.\
  Commun.\  \textbf{178} (2008) 852},
  \href{http://arxiv.org/abs/0710.3820}{{\normalfont\ttfamily
  arXiv:0710.3820}}\relax
\mciteBstWouldAddEndPuncttrue
\mciteSetBstMidEndSepPunct{\mcitedefaultmidpunct}
{\mcitedefaultendpunct}{\mcitedefaultseppunct}\relax
\EndOfBibitem
\bibitem{LHCb-PROC-2010-056}
I.~Belyaev {\em et~al.}, \ifthenelse{\boolean{articletitles}}{\emph{{Handling
  of the generation of primary events in Gauss, the LHCb simulation
  framework}}, }{}\href{https://doi.org/10.1088/1742-6596/331/3/032047}{J.\
  Phys.\ Conf.\ Ser.\  \textbf{331} (2011) 032047}\relax
\mciteBstWouldAddEndPuncttrue
\mciteSetBstMidEndSepPunct{\mcitedefaultmidpunct}
{\mcitedefaultendpunct}{\mcitedefaultseppunct}\relax
\EndOfBibitem
\bibitem{Lange:2001uf}
D.~J. Lange, \ifthenelse{\boolean{articletitles}}{\emph{{The EvtGen particle
  decay simulation package}},
  }{}\href{https://doi.org/10.1016/S0168-9002(01)00089-4}{Nucl.\ Instrum.\
  Meth.\  \textbf{A462} (2001) 152}\relax
\mciteBstWouldAddEndPuncttrue
\mciteSetBstMidEndSepPunct{\mcitedefaultmidpunct}
{\mcitedefaultendpunct}{\mcitedefaultseppunct}\relax
\EndOfBibitem
\bibitem{davidson2015photos}
N.~Davidson, T.~Przedzinski, and Z.~Was,
  \ifthenelse{\boolean{articletitles}}{\emph{{PHOTOS interface in C++:
  Technical and physics documentation}},
  }{}\href{https://doi.org/https://doi.org/10.1016/j.cpc.2015.09.013}{Comp.\
  Phys.\ Comm.\  \textbf{199} (2016) 86},
  \href{http://arxiv.org/abs/1011.0937}{{\normalfont\ttfamily
  arXiv:1011.0937}}\relax
\mciteBstWouldAddEndPuncttrue
\mciteSetBstMidEndSepPunct{\mcitedefaultmidpunct}
{\mcitedefaultendpunct}{\mcitedefaultseppunct}\relax
\EndOfBibitem
\bibitem{Allison:2006ve}
Geant4 collaboration, J.~Allison {\em et~al.},
  \ifthenelse{\boolean{articletitles}}{\emph{{Geant4 developments and
  applications}}, }{}\href{https://doi.org/10.1109/TNS.2006.869826}{IEEE
  Trans.\ Nucl.\ Sci.\  \textbf{53} (2006) 270}\relax
\mciteBstWouldAddEndPuncttrue
\mciteSetBstMidEndSepPunct{\mcitedefaultmidpunct}
{\mcitedefaultendpunct}{\mcitedefaultseppunct}\relax
\EndOfBibitem
\bibitem{LHCb-PROC-2011-006}
M.~Clemencic {\em et~al.}, \ifthenelse{\boolean{articletitles}}{\emph{{The
  \lhcb simulation application, Gauss: Design, evolution and experience}},
  }{}\href{https://doi.org/10.1088/1742-6596/331/3/032023}{J.\ Phys.\ Conf.\
  Ser.\  \textbf{331} (2011) 032023}\relax
\mciteBstWouldAddEndPuncttrue
\mciteSetBstMidEndSepPunct{\mcitedefaultmidpunct}
{\mcitedefaultendpunct}{\mcitedefaultseppunct}\relax
\EndOfBibitem
\bibitem{LHCb-PAPER-2022-045}
LHCb collaboration, R.~Aaij {\em et~al.},
  \ifthenelse{\boolean{articletitles}}{\emph{{Measurement of lepton
  universality parameters in $\Bp\to\Kp\ellp\ellm$ and
  $\Bz\to\Kstarz\ellp\ellm$ decays}},
  }{}\href{https://doi.org/10.1103/PhysRevD.108.032002}{Phys.\ Rev.\
  \textbf{D108} (2023) 032002},
  \href{http://arxiv.org/abs/2212.09153}{{\normalfont\ttfamily
  arXiv:2212.09153}}\relax
\mciteBstWouldAddEndPuncttrue
\mciteSetBstMidEndSepPunct{\mcitedefaultmidpunct}
{\mcitedefaultendpunct}{\mcitedefaultseppunct}\relax
\EndOfBibitem
\bibitem{PDG2024}
Particle Data Group, S.~Navas {\em et~al.},
  \ifthenelse{\boolean{articletitles}}{\emph{{\href{http://pdg.lbl.gov/}{Review
  of particle physics}}},
  }{}\href{https://doi.org/10.1103/PhysRevD.110.030001}{Phys.\ Rev.\
  \textbf{D110} (2024) 030001}\relax
\mciteBstWouldAddEndPuncttrue
\mciteSetBstMidEndSepPunct{\mcitedefaultmidpunct}
{\mcitedefaultendpunct}{\mcitedefaultseppunct}\relax
\EndOfBibitem
\bibitem{Hulsbergen:2005pu}
W.~D. Hulsbergen, \ifthenelse{\boolean{articletitles}}{\emph{{Decay chain
  fitting with a Kalman filter}},
  }{}\href{https://doi.org/10.1016/j.nima.2005.06.078}{Nucl.\ Instrum.\ Meth.\
  \textbf{A552} (2005) 566},
  \href{http://arxiv.org/abs/physics/0503191}{{\normalfont\ttfamily
  arXiv:physics/0503191}}\relax
\mciteBstWouldAddEndPuncttrue
\mciteSetBstMidEndSepPunct{\mcitedefaultmidpunct}
{\mcitedefaultendpunct}{\mcitedefaultseppunct}\relax
\EndOfBibitem
\bibitem{XGBoost}
T.~Chen and C.~Guestrin, \ifthenelse{\boolean{articletitles}}{\emph{{XGBoost: a
  scalable tree boosting system}}, }{} in {\em Proceedings of the 22nd ACM
  SIGKDD International Conference on Knowledge Discovery and Data Mining},
  \href{https://doi.org/10.1145/2939672.2939785}{ KDD '16, (New York, NY, USA),
  785–794, Association for Computing Machinery, 2016},
  \href{http://arxiv.org/abs/1603.02754}{{\normalfont\ttfamily
  arXiv:1603.02754}}\relax
\mciteBstWouldAddEndPuncttrue
\mciteSetBstMidEndSepPunct{\mcitedefaultmidpunct}
{\mcitedefaultendpunct}{\mcitedefaultseppunct}\relax
\EndOfBibitem
\bibitem{10.1145/307400.307439}
A.~Blum, A.~Kalai, and J.~Langford,
  \ifthenelse{\boolean{articletitles}}{\emph{Beating the hold-out: bounds for
  k-fold and progressive cross-validation}, }{} in {\em Proceedings of the
  Twelfth Annual Conference on Computational Learning Theory},
  \href{https://doi.org/10.1145/307400.307439}{ COLT '99, (New York, NY, USA),
  203–208, Association for Computing Machinery, 1999}\relax
\mciteBstWouldAddEndPuncttrue
\mciteSetBstMidEndSepPunct{\mcitedefaultmidpunct}
{\mcitedefaultendpunct}{\mcitedefaultseppunct}\relax
\EndOfBibitem
\bibitem{dalitz_decay}
R.~H. Dalitz, \ifthenelse{\boolean{articletitles}}{\emph{On an alternative
  decay process for the neutral $\pi$-meson},
  }{}\href{https://doi.org/10.1088/0370-1298/64/7/115}{Proceedings of the
  Physical Society.\ Section A \textbf{64} (1951) 667}\relax
\mciteBstWouldAddEndPuncttrue
\mciteSetBstMidEndSepPunct{\mcitedefaultmidpunct}
{\mcitedefaultendpunct}{\mcitedefaultseppunct}\relax
\EndOfBibitem
\bibitem{Skwarnicki:1986xj}
T.~Skwarnicki, {\em {A study of the radiative cascade transitions between the
  Upsilon-prime and Upsilon resonances}}, PhD thesis, Institute of Nuclear
  Physics, Krakow, 1986,
  {\href{http://inspirehep.net/record/230779/}{DESY-F31-86-02}}\relax
\mciteBstWouldAddEndPuncttrue
\mciteSetBstMidEndSepPunct{\mcitedefaultmidpunct}
{\mcitedefaultendpunct}{\mcitedefaultseppunct}\relax
\EndOfBibitem
\bibitem{Cranmer2001}
K.~Cranmer, \ifthenelse{\boolean{articletitles}}{\emph{Kernel estimation in
  high-energy physics},
  }{}\href{https://doi.org/10.1016/s0010-4655(00)00243-5}{Computer Physics
  Communications \textbf{136} (2001) 198–207},
  \href{http://arxiv.org/abs/hep-ex/0011057}{{\normalfont\ttfamily
  arXiv:hep-ex/0011057}}\relax
\mciteBstWouldAddEndPuncttrue
\mciteSetBstMidEndSepPunct{\mcitedefaultmidpunct}
{\mcitedefaultendpunct}{\mcitedefaultseppunct}\relax
\EndOfBibitem
\bibitem{bib:Feldman-Cousins}
G.~J. Feldman and R.~D. Cousins,
  \ifthenelse{\boolean{articletitles}}{\emph{Unified approach to the classical
  statistical analysis of small signals},
  }{}\href{https://doi.org/10.1103/PhysRevD.57.3873}{Phys.\ Rev.\  \textbf{D57}
  (1998) 3873},
  \href{http://arxiv.org/abs/physics/9711021}{{\normalfont\ttfamily
  arXiv:physics/9711021}}\relax
\mciteBstWouldAddEndPuncttrue
\mciteSetBstMidEndSepPunct{\mcitedefaultmidpunct}
{\mcitedefaultendpunct}{\mcitedefaultseppunct}\relax
\EndOfBibitem
\bibitem{LHCb-PAPER-2024-032}
LHCb collaboration, R.~Aaij {\em et~al.},
  \ifthenelse{\boolean{articletitles}}{\emph{{Test of lepton flavour
  universality with $B_s^0 \rightarrow \phi \ell^+\ell^-$ decays}},
  }{}\href{http://arxiv.org/abs/2410.13748}{{\normalfont\ttfamily
  arXiv:2410.13748}}, {Submitted to Phys. Rev. Lett.}\relax
\mciteBstWouldAddEndPunctfalse
\mciteSetBstMidEndSepPunct{\mcitedefaultmidpunct}
{}{\mcitedefaultseppunct}\relax
\EndOfBibitem
\bibitem{Be_irevi__2012}
D.~Bečirević and E.~Schneider, \ifthenelse{\boolean{articletitles}}{\emph{{On
  transverse asymmetries in $B \to K^{*} \ell^+\ell^-$}},
  }{}\href{https://doi.org/10.1016/j.nuclphysb.2011.09.004}{Nucl.\ Phys.\
  \textbf{B854} (2012) 321–339},
  \href{http://arxiv.org/abs/1106.3283}{{\normalfont\ttfamily
  arXiv:1106.3283}}\relax
\mciteBstWouldAddEndPuncttrue
\mciteSetBstMidEndSepPunct{\mcitedefaultmidpunct}
{\mcitedefaultendpunct}{\mcitedefaultseppunct}\relax
\EndOfBibitem
\bibitem{straub2018flavio}
D.~M. Straub, \ifthenelse{\boolean{articletitles}}{\emph{{flavio: a Python
  package for flavour and precision phenomenology in the Standard Model and
  beyond}}, }{}\href{http://arxiv.org/abs/1810.08132}{{\normalfont\ttfamily
  arXiv:1810.08132}}\relax
\mciteBstWouldAddEndPuncttrue
\mciteSetBstMidEndSepPunct{\mcitedefaultmidpunct}
{\mcitedefaultendpunct}{\mcitedefaultseppunct}\relax
\EndOfBibitem
\bibitem{Jager:2014rwa}
S.~J\"ager and J.~Martin~Camalich,
  \ifthenelse{\boolean{articletitles}}{\emph{{{Reassessing the discovery
  potential of the $B \to K^{*} \ell^+\ell^-$ decays in the large-recoil
  region: SM challenges and BSM opportunities}}},
  }{}\href{https://doi.org/10.1103/PhysRevD.93.014028}{Phys.\ Rev.\
  \textbf{D93} (2016) 014028},
  \href{http://arxiv.org/abs/1412.3183}{{\normalfont\ttfamily
  arXiv:1412.3183}}\relax
\mciteBstWouldAddEndPuncttrue
\mciteSetBstMidEndSepPunct{\mcitedefaultmidpunct}
{\mcitedefaultendpunct}{\mcitedefaultseppunct}\relax
\EndOfBibitem
\bibitem{Paul2017}
A.~Paul and D.~M. Straub,
  \ifthenelse{\boolean{articletitles}}{\emph{{Constraints on new physics from
  radiative $B$ decays}},
  }{}\href{https://doi.org/10.1007/jhep04(2017)027}{JHEP \textbf{2017} (2017)
  }, \href{http://arxiv.org/abs/1608.02556}{{\normalfont\ttfamily
  arXiv:1608.02556}}\relax
\mciteBstWouldAddEndPuncttrue
\mciteSetBstMidEndSepPunct{\mcitedefaultmidpunct}
{\mcitedefaultendpunct}{\mcitedefaultseppunct}\relax
\EndOfBibitem
\end{mcitethebibliography}

\newpage
\centerline
{\large\bf LHCb collaboration}
\begin
{flushleft}
\small
R.~Aaij$^{38}$\lhcborcid{0000-0003-0533-1952},
A.S.W.~Abdelmotteleb$^{57}$\lhcborcid{0000-0001-7905-0542},
C.~Abellan~Beteta$^{51}$,
F.~Abudin{\'e}n$^{57}$\lhcborcid{0000-0002-6737-3528},
T.~Ackernley$^{61}$\lhcborcid{0000-0002-5951-3498},
A. A. ~Adefisoye$^{69}$\lhcborcid{0000-0003-2448-1550},
B.~Adeva$^{47}$\lhcborcid{0000-0001-9756-3712},
M.~Adinolfi$^{55}$\lhcborcid{0000-0002-1326-1264},
P.~Adlarson$^{82}$\lhcborcid{0000-0001-6280-3851},
C.~Agapopoulou$^{14}$\lhcborcid{0000-0002-2368-0147}
% ...

{\footnotesize \it

% ...
$^{14}$Universit{\'e} Paris-Saclay, CNRS/IN2P3, IJCLab, Orsay, France\\
% ...
$^{38}$Nikhef National Institute for Subatomic Physics, Amsterdam, Netherlands\\
% ...
$^{47}$Instituto Galego de F{\'\i}sica de Altas Enerx{\'\i}as (IGFAE), Universidade de Santiago de Compostela, Santiago de Compostela, Spain\\
% ...
$^{51}$Physik-Institut, Universit{\"a}t Z{\"u}rich, Z{\"u}rich, Switzerland\\
% ...
$^{55}$H.H. Wills Physics Laboratory, University of Bristol, Bristol, United Kingdom\\
% ...
$^{57}$Department of Physics, University of Warwick, Coventry, United Kingdom\\
% ...
$^{60}$School of Physics and Astronomy, University of Glasgow, Glasgow, United Kingdom\\
$^{61}$Oliver Lodge Laboratory, University of Liverpool, Liverpool, United Kingdom\\
% ...
$^{69}$Syracuse University, Syracuse, NY, United States\\
% ...
$^{82}$Department of Physics and Astronomy, Uppsala University, Uppsala, Sweden, associated to $^{60}$\\
% ...
}
\end{flushleft}

\end{document}